\documentclass[preprint,10pt,authoryear]{elsarticle}
\usepackage[utf8]{inputenc}
\usepackage[T1]{fontenc}

\usepackage[a4paper,top=3.6cm,bottom=2.2cm,left=2.3cm,right=2.3cm]{geometry}

\usepackage{amsmath,amssymb,amsfonts}
\usepackage{bm}

\usepackage{graphicx}
\usepackage{booktabs}
\usepackage{threeparttable}
\usepackage{longtable}
\usepackage{array}
\usepackage{multirow}
\usepackage{caption}
\usepackage{subcaption}
\usepackage{float}
\usepackage{placeins}
\usepackage{tabularx}
\usepackage{pdflscape}
\usepackage{xcolor}

\usepackage{microtype}

\usepackage{enumitem}
\setlist[itemize]{topsep=3pt,itemsep=2pt}
\setlist[enumerate]{topsep=3pt,itemsep=2pt}

\usepackage{ragged2e}
\newcolumntype{Y}{>{\RaggedRight\arraybackslash}X}
\newcolumntype{P}[1]{>{\RaggedRight\arraybackslash}p{#1}}
\newcolumntype{Q}[1]{>{\Centering\arraybackslash}p{#1}}

\newcommand{\figurenote}[1]{%
\vspace{0.3em}
\begin{minipage}{0.90\textwidth}
\footnotesize
\textit{Note:} #1
\end{minipage}
}
\usepackage{caption}

\newcommand{\tablenote}[1]{%
\vspace{0.3em}
\begin{minipage}{0.95\textwidth}
\footnotesize
\textit{Note:} #1
\end{minipage}
}

\usepackage{natbib}
\usepackage{hyperref}
\hypersetup{
    hidelinks,
    pdftitle={Volatility Forecasting and Return Prediction under Market Regimes},
    pdfauthor={Xinyue Fang and Robert Ślepaczuk}
}

\journal{Preprint}

\begin{document}

\begin{frontmatter}

\title{Volatility Forecasting and Return Prediction under Market Regimes: Evidence from High-Frequency Chinese Equity Data}

\author[a]{Xinyue Fang\fnref{fn1}}

\author[b]{Robert Ślepaczuk\corref{cor1}\fnref{fn2}}

\address[a]{Quantitative Finance Research Group, Faculty of Economic Sciences, University of Warsaw, ul. Długa 44/50, 00-241 Warsaw, Poland}

\address[b]{Quantitative Finance Research Group, Department of Quantitative Finance and Machine Learning, Faculty of Economic Sciences, University of Warsaw, ul. Długa 44/50, 00-241 Warsaw, Poland}

\cortext[cor1]{Corresponding author. Email address: rslepaczuk@wne.uw.edu.pl}

\fntext[fn1]{ORCID: 0009-0007-4538-4668}
\fntext[fn2]{ORCID: 0000-0001-5227-2014}

\begin{abstract}

This study investigates whether regime-dependent volatility forecasting and machine-learning-based return prediction can be jointly integrated to improve both statistical forecasting performance and economic strategy outcomes in equity markets. Using high-frequency CSI 300 Index data from 2005 to 2023, a sequential two-stage framework is developed. In the first stage, realized volatility is modeled using regime-augmented HARQ specifications combined with Markov-switching GJR-GARCH filtering to capture long-memory dynamics, asymmetry, and structural market regimes. In the second stage, volatility forecasts, regime indicators, and return-related predictors are incorporated into an XGBoost return-prediction model estimated through a strictly walk-forward out-of-sample procedure. The empirical results demonstrate that regime-aware volatility forecasting consistently outperforms baseline HARQ models across forecast evaluation metrics and is generally supported by formal forecast comparison tests. In contrast, return predictability remains weak, state-dependent, and concentrated primarily in low-volatility regimes. Although naive predictive trading strategies generally fail after accounting for realistic transaction costs, carefully designed implementations incorporating volatility scaling, low-volatility gating, threshold calibration, and turnover controls can improve defensive economic performance. The findings suggest that the practical value of predictive systems in financial markets may depend less on generating strong unconditional return forecasts and more on transforming weak state-dependent signals into economically robust portfolio allocation rules. Overall, the study contributes by integrating econometric volatility modeling, regime classification, machine-learning return prediction, and implementation realism within a unified framework.

\end{abstract}


\end{frontmatter}

\setlength\parindent{1cm}

\section{Introduction}
\label{sec:introduction}

Forecasting financial market dynamics and designing economically viable investment strategies remain central challenges in empirical finance. A large body of literature has established that volatility exhibits substantial persistence and predictability, particularly when measured using high-frequency-based realized measures \citep{andersen2003,corsi2009}. In contrast, return predictability is generally weak, unstable, and highly sensitive to model specification, reflecting the low signal-to-noise ratio inherent in financial returns. This creates a fundamental tension in financial forecasting: although risk can often be forecasted with reasonable accuracy, translating predictive information into economically meaningful trading performance remains considerably more difficult. Understanding how volatility dynamics, market regimes, and return predictability interact is therefore of both academic and practical importance.

Two major strands of literature have evolved to address these challenges. The first focuses on volatility forecasting. Realized-volatility-based models, particularly HAR-type specifications, have become standard tools for capturing the heterogeneous and persistent structure of financial volatility \citep{corsi2009,bollerslev2016}. More recently, researchers have incorporated regime-switching mechanisms into volatility models, allowing volatility dynamics to differ across market states and improving forecasting performance during periods of structural change \citep{hamilton1989,klaassen2002,ma2019}. The second strand focuses on return prediction. Machine learning methods have attracted increasing attention because of their ability to capture nonlinear relationships, complex interactions, and high-dimensional predictor sets \citep{chen2016,feng2026}. Hybrid approaches that combine econometric structures with machine learning techniques have also emerged, reflecting the potential benefits of integrating economic interpretability with predictive flexibility \citep{benromdhane2026,peter2026}.

Despite these advances, the literature remains fragmented. Volatility forecasting and return prediction are frequently treated as separate problems, while relatively limited attention has been paid to how regime-dependent volatility information can be systematically incorporated into machine-learning-based return forecasting frameworks. Moreover, even when statistical predictability is identified, its economic value depends critically on implementation choices, including portfolio construction, turnover control, transaction costs, and model re-estimation procedures. Recent research increasingly emphasizes that statistically significant forecasts do not necessarily translate into economically significant investment performance, highlighting the importance of realistic out-of-sample evaluation and implementation-aware strategy design \citep{gu2020}.

This study seeks to bridge these gaps by developing an integrated framework that combines regime-dependent volatility forecasting, machine-learning-based return prediction, and realistic strategy implementation. Using high-frequency CSI 300 Index data from the Chinese equity market over the period 2005--2023, the analysis proceeds in two stages. First, realized volatility is modeled using regime-augmented HARQ specifications combined with Markov-switching GJR-GARCH filtering to capture persistence, asymmetry, and structural market regimes. Second, volatility forecasts, regime variables, and return-related predictors are incorporated into an XGBoost return-prediction framework estimated through a strictly walk-forward out-of-sample procedure. The resulting forecasts are subsequently evaluated through a range of trading-strategy implementations that account for volatility scaling, threshold calibration, turnover control, and transaction costs.

The empirical findings reveal a clear contrast between volatility forecasting and return prediction. Regime-aware volatility models consistently outperform baseline HARQ specifications across multiple forecast evaluation metrics and statistical tests, confirming the importance of incorporating regime information into volatility dynamics. By contrast, return predictability remains weak and highly state-dependent, with economically relevant predictive content concentrated primarily in low-volatility regimes. Naive predictive strategies generally fail after realistic trading frictions are considered. However, carefully engineered implementations that combine low-volatility gating, walk-forward calibration, risk scaling, and turnover controls can improve defensive portfolio performance and downside-risk characteristics, although such gains remain sensitive to implementation choices and are not uniformly robust across alternative specifications.

The analysis contributes to the literature in several respects. First, it develops an integrated framework linking regime-dependent volatility forecasting and machine-learning-based return prediction, thereby connecting two strands of research that are often studied separately. Second, it provides a comprehensive evaluation that jointly considers statistical forecasting accuracy, economic performance, and implementation constraints within a unified out-of-sample setting. Third, it highlights the role of volatility and regime information as economically valuable risk-management signals, even when directional return predictability remains weak. Fourth, it demonstrates how implementation design, including signal engineering, volatility scaling, and turnover management, can materially influence the economic value extracted from predictive models.

The remainder of the paper is organized as follows. Section~\ref{sec:literature} reviews the related literature. Section~\ref{sec:data} describes the data and empirical design. Section~\ref{sec:methodology} presents the econometric and machine-learning methodology. Section~\ref{sec:results} reports the empirical findings. Section~\ref{sec:robustness} provides robustness analyses and diagnostic tests. Section~\ref{sec:conclusion} concludes.

\section{Related literature}
\label{sec:literature}

\subsection{Realized measures of volatility}

Accurate measurement of financial market volatility is fundamental to risk management, asset pricing, and empirical forecasting. Traditional approaches typically rely on latent volatility models, where volatility is unobserved and must be inferred indirectly from low-frequency returns. The availability of high-frequency data has led to a major advancement in this area through the development of realized measures.

The concept of realized variance, introduced by \citet{andersen1998} and further developed by \citet{andersen2001,andersen2003}, provides a nonparametric estimator of the ex-post variation of asset prices. It is defined as:
\begin{equation}
RV_t = \sum_{i=1}^{M} r_{t,i}^2,
\label{eq:rv}
\end{equation}
where $r_{t,i}$ denotes the intraday return over interval $i$ within day $t$, and $M$ is the number of intraday observations. Under ideal conditions, $RV_t$ provides a consistent estimator of the integrated variance of the underlying price process \citep{andersen2003}.

Compared with traditional low-frequency volatility proxies, realized variance uses high-frequency information directly and therefore provides a more informative measure of ex-post market variability. This makes it particularly suitable for empirical volatility forecasting.

To capture higher-order variation, realized quarticity is defined as:
\begin{equation}
RQ_t = \frac{M}{3} \sum_{i=1}^{M} r_{t,i}^4,
\label{eq:rq}
\end{equation}
which serves as a proxy for the variability of realized variance itself. \citet{bollerslev2016} show that incorporating realized quarticity can improve volatility forecasts by capturing time variation in volatility-of-volatility.

Another important extension is the decomposition of volatility into continuous and jump components. \citet{barndorff2004} introduce bipower variation as a robust estimator of the continuous sample-path component. The discontinuous jump component can then be approximated as:
\begin{equation}
CJ_t = \max(RV_t - BPV_t, 0),
\label{eq:cj}
\end{equation}
where $BPV_t$ denotes bipower variation.

The literature also emphasizes that jumps may have asymmetric effects. Using a semivariance-based framework, \citet{patton2015} distinguish between positive and negative jump contributions. Following this intuition, this study employs a parsimonious signed-jump proxy:
\begin{equation}
\text{SignedJump}_t = CJ_t \cdot \mathrm{sign}(r_t),
\label{eq:signedjump}
\end{equation}
which captures both jump magnitude and market direction.

Overall, realized variance, realized quarticity, and jump-related measures provide a rich set of inputs for volatility modeling. As noted by \citet{leushuis2026}, high-frequency-based realized measures have become standard tools in modern volatility forecasting.

\subsection{HARQ-based volatility modeling}

While realized measures provide accurate estimates of ex-post volatility, they exhibit strong persistence and long-memory behavior. To model this feature, \citet{corsi2009} propose the Heterogeneous Autoregressive model:
\begin{equation}
RV_t = \beta_0 + \beta_1 RV_{t-1} + \beta_2 RV_{t-5:t-1} + \beta_3 RV_{t-22:t-1} + \varepsilon_t.
\label{eq:har}
\end{equation}

The HAR model captures long memory through volatility components defined over heterogeneous time horizons, reflecting the behavior of market participants operating at different frequencies. Despite its simplicity, the HAR model has proven robust across markets and time periods, making it a benchmark specification in realized-volatility forecasting.

To further enhance predictive performance, \citet{bollerslev2016} extend the HAR model by incorporating realized quarticity:
\begin{equation}
RV_t = \beta_0 + \beta_1 RV_{t-1} + \beta_2 RV_{t-5:t-1} + \beta_3 RV_{t-22:t-1} + \beta_4 RQ_{t-1} + \varepsilon_t.
\label{eq:harq}
\end{equation}

In practice, realized volatility measures are often highly skewed and heavy-tailed. To address this issue, the HARQ model is commonly estimated in logarithmic form:
\begin{equation}
\log(RV_t) = \beta_0 + \beta_1 \log(RV_{t-1}) + \beta_2 \log(RV_{t-5:t-1}) + \beta_3 \log(RV_{t-22:t-1}) + \beta_4 \log(RQ_{t-1}) + \varepsilon_t.
\label{eq:logharq}
\end{equation}

This transformation improves statistical properties and stabilizes variance, thereby enhancing the suitability of linear modeling. In the present study, the HARQ framework provides the baseline volatility forecasts and the foundation for incorporating regime-dependent dynamics.

\subsection{Regime-switching in volatility dynamics}

Although HAR-type models capture volatility persistence effectively, they generally assume stable relationships over time. In practice, financial markets frequently alternate between distinct states, such as low-volatility and high-volatility periods. To account for such structural changes, regime-switching models have been widely adopted.

The Markov-switching framework introduced by \citet{hamilton1989} and further developed by \citet{hamilton1994} allows model parameters to depend on an unobserved state variable that evolves according to a Markov process. This framework provides a flexible way to capture nonlinear and state-dependent dynamics.

In volatility modeling, \citet{klaassen2002} show that allowing GARCH dynamics to vary across regimes can improve forecasting performance. Within the GARCH family, asymmetry is captured by the GJR-GARCH model \citep{glosten1993}:
\begin{equation}
\sigma_t^2 = \omega + \alpha \varepsilon_{t-1}^2 + \gamma \varepsilon_{t-1}^2 I(\varepsilon_{t-1}<0) + \beta \sigma_{t-1}^2,
\label{eq:gjrgarch}
\end{equation}
while heavy-tailed innovations are commonly modeled using Student-$t$ distributions \citep{bollerslev1987}.

Recent studies extend these ideas to realized-volatility forecasting by incorporating regime dependence into HAR-type frameworks, particularly in Chinese financial markets. \citet{wang2019} develop a Markov-switching HAR model with time-varying transition probabilities for the Shanghai Stock Exchange Composite Index and show that regime-switching specifications consistently outperform standard HAR benchmarks. Although their study focuses on a different Chinese equity index and emphasizes transition asymmetries, leverage effects, and trading-volume asymmetries rather than the CSI 300 specifically, their findings provide evidence that realized-volatility dynamics in Chinese equity markets are materially asymmetric and regime dependent.

\citet{ma2019} focus on CSI 300 index futures rather than the CSI 300 spot index itself and propose an MS-HARQ-FIGARCH framework that integrates long memory, regime shifts, and GARCH-family volatility dynamics. Their results show that adding regime-switching and GARCH-family structures improves volatility forecasting performance relative to simpler benchmarks. While the underlying asset differs from the present study, the central conclusion is directionally consistent: incorporating regime dependence and richer conditional variance dynamics can enhance predictive accuracy in Chinese-market volatility forecasting.

The present study extends this literature by focusing on high-frequency CSI 300 spot-index realized volatility over a long sample period and by integrating regime-dependent volatility forecasts into a downstream return-prediction framework.

Building on this literature, the present study adopts a sequential two-stage design. HARQ residuals are first modeled using a Markov-switching GJR-GARCH process to obtain filtered regime probabilities, which are subsequently incorporated as additional explanatory variables in regime-augmented HARQ specifications.This sequential design combines the heterogeneous long-memory structure of HARQ with regime-dependent conditional variance and asymmetric shock responses. Relative to prior studies, the framework provides an integrated extension that applies asymmetric regime-switching volatility modeling to CSI 300 spot-index realized volatility, uses filtered regime probabilities as predictive state variables, and evaluates whether volatility-regime information improves both volatility forecasting and economically implementable return prediction.

\subsection{Machine learning and hybrid approaches to volatility forecasting}

While econometric models offer strong interpretability, they may be limited in capturing complex nonlinear patterns. This has motivated increasing interest in integrating machine learning techniques into volatility forecasting.

Recent studies propose hybrid models that combine econometric and neural-network structures. For example, \citet{benromdhane2026} develop a HAR-LSTM-GARCH framework, \citet{peter2026} propose a GARCH-LSTM stacking model, and \citet{sukainah2026} combine GARCH models with neural networks. Other studies apply deep learning more directly. \citet{jain2026} explore deep learning and reinforcement learning for volatility forecasting and portfolio optimization, while \citet{zhang2026} incorporate sentiment information into deep-learning-based volatility models. In addition, \citet{yihuan2026} embed GARCH structures within neural-network architectures.

These approaches highlight the potential of combining econometric structure with machine-learning flexibility, although they often involve trade-offs between interpretability and predictive power.

\subsection{Machine learning for return prediction}

Return prediction remains particularly challenging because of the low signal-to-noise ratio in financial data. Linear models often fail to capture nonlinear interactions and complex dependencies among predictors. Machine learning methods, especially gradient boosting techniques such as XGBoost \citep{chen2016}, provide a flexible framework for addressing these challenges. By constructing an ensemble of decision trees and optimizing a regularized objective function, XGBoost can capture nonlinear relationships while controlling overfitting.

Empirical applications demonstrate the growing relevance of machine learning in financial prediction. \citet{feng2026} compare LSTM and XGBoost in stock-price prediction, while \citet{li2026} apply XGBoost to mutual-fund return forecasting.

An important consideration in financial machine learning is evaluation methodology. Due to the time-series structure and nonstationary nature of financial data, random train--test splits are generally inappropriate because they risk look-ahead bias and unrealistic information sets. Instead, rolling, expanding-window, or walk-forward validation frameworks are widely used because they better mimic real-time forecasting by sequentially updating the estimation sample. In large-scale empirical asset-pricing applications, \citet{gu2020} emphasize rigorous out-of-sample evaluation and careful control of overfitting when applying machine-learning models to financial prediction problems.

This approach provides a more realistic assessment of predictive performance and helps mitigate overfitting in nonstationary environments.

\subsection{Research gap and contribution}

Despite extensive research, volatility modeling, regime-switching dynamics, and machine-learning prediction are often studied separately. Volatility models typically focus on statistical forecasting accuracy, while machine-learning approaches emphasize flexible prediction but often provide limited economic interpretation.

This study contributes by integrating these approaches within a unified empirical framework. Realized measures are used to construct volatility inputs, HARQ models capture long-memory dynamics, and Markov-switching GJR-GARCH models account for state-dependent conditional variance and asymmetric shock responses. These components are then combined with machine-learning techniques for return prediction under a strictly walk-forward evaluation framework.

This integrated approach allows both statistical and economic aspects of predictability to be examined within a consistent empirical setting. It also provides a bridge between volatility forecasting, return prediction, and trading-strategy implementation, thereby addressing the practical question of whether regime-dependent volatility information can be transformed into economically meaningful portfolio decisions.

\section{Data and empirical design}
\label{sec:data}

\subsection{Data}

This study uses high-frequency data for the CSI 300 Index (sh000300), which is published by the Shanghai Stock Exchange and represents a broad measure of the Chinese A-share market. The data consist of 5-minute intraday price observations obtained from the Wind Financial Terminal provided by Wind Information Co., Ltd. The sample period spans from April 8, 2005 to May 31, 2023.

Given the trading mechanism of the Chinese A-share market, specific data-filtering procedures are applied before constructing realized measures. To avoid distortions arising from overnight returns, opening-auction effects, and non-trading intervals, observations before 09:30 and during the midday break from 11:30 to 13:00 are excluded. The analysis therefore retains only the continuous trading sessions between 09:30--11:30 and 13:00--15:00.

The first 5-minute return after market opening is constructed exclusively from prices within the continuous trading session rather than using the previous day's closing price. Consequently, overnight price movements are excluded from intraday realized-variance calculations. Similarly, the midday break is not treated as a continuous trading interval. This filtering procedure ensures that realized measures are based solely on actively traded periods and are not mechanically inflated by overnight or non-trading-period discontinuities.

No additional noise-robust realized-volatility estimators, such as two-scale realized variance or realized kernel estimators, are employed in the baseline analysis. Instead, the study follows the common empirical convention of using 5-minute sampling frequencies, which are widely regarded as a practical compromise between reducing market microstructure noise and preserving sufficient intraday variation for volatility measurement \citep{andersen2001,barndorff2004,andersen2011}. Nevertheless, microstructure noise may still affect realized measures, particularly around market open and close periods and during earlier segments of the sample. This limitation is acknowledged when interpreting the empirical results.

Daily returns are computed as logarithmic differences of closing prices:
\begin{equation}
r_t = \log(P_t)-\log(P_{t-1}),
\end{equation}
where $P_t$ denotes the closing price on day $t$.

Intraday returns are used to construct realized variance:
\begin{equation}
RV_t = \sum_{i=1}^{M_t} r_{t,i}^{2},
\end{equation}
where $r_{t,i}$ denotes the intraday return for interval $i$ on day $t$, and $M_t$ denotes the number of intraday observations.

In addition, realized quarticity ($RQ_t$) and jump variation ($CJ_t$) are constructed to capture higher-order volatility dynamics and discontinuous price movements. These realized measures are derived from the same filtered 5-minute return series to ensure consistency across volatility-related predictors.

To stabilize the distribution of volatility, the logarithm of realized variance is used throughout the volatility-modeling stage. The transformed variable $\log(RV_t)$ serves as the primary dependent variable in the forecasting framework.

All predictors are constructed using information available up to time $t$, whereas the target variable in the return-prediction model is the return at time $t+1$. This timing convention ensures a strictly out-of-sample forecasting environment and avoids the use of future information.

\subsection{Descriptive statistics}

Table~\ref{tab:dist_stats} reports summary statistics for daily returns, realized variance ($RV_t$), and log-transformed realized variance ($\log(RV_t)$).

\begin{table}[H]
\centering
\footnotesize
\setlength{\tabcolsep}{4pt}
\renewcommand{\arraystretch}{1.05}
\caption{Distribution summary statistics}
\label{tab:dist_stats}

\begin{tabular}{lcccccc}
\toprule
Variable & Mean & Std & Skewness & Kurtosis & JB Stat & JB p-value \\
\midrule
Return   & 0.0003 & 0.0167 & -0.5652 & 7.4221 & 3758.48 & 0.0000 \\
RV       & 0.0002 & 0.0003 & 6.1377  & 61.3227 & 641339.63 & 0.0000 \\
log(RV)  & -9.1938 & 1.0076 & 0.4815 & 3.1289 & 170.61 & 0.0000 \\
\bottomrule
\end{tabular}

\vspace{0.4em}

\begin{minipage}{0.90\textwidth}
\footnotesize
\textit{Note:} The table reports summary statistics for daily returns, realized variance ($RV_t$), and log-transformed realized variance ($\log(RV_t)$). JB denotes the Jarque--Bera test statistic for normality.
\end{minipage}

\end{table}

The distribution of daily returns exhibits substantial deviations from normality. Returns are negatively skewed and display excess kurtosis, indicating asymmetric and heavy-tailed behavior. The Jarque--Bera test strongly rejects the null hypothesis of normality. These findings are consistent with well-documented stylized facts of financial returns and support the use of heavy-tailed distributions in subsequent volatility modeling.

\begin{figure}[H]
\centering
\caption{Distribution of daily returns}
\label{fig:return_hist}
\includegraphics[width=0.60\textwidth]{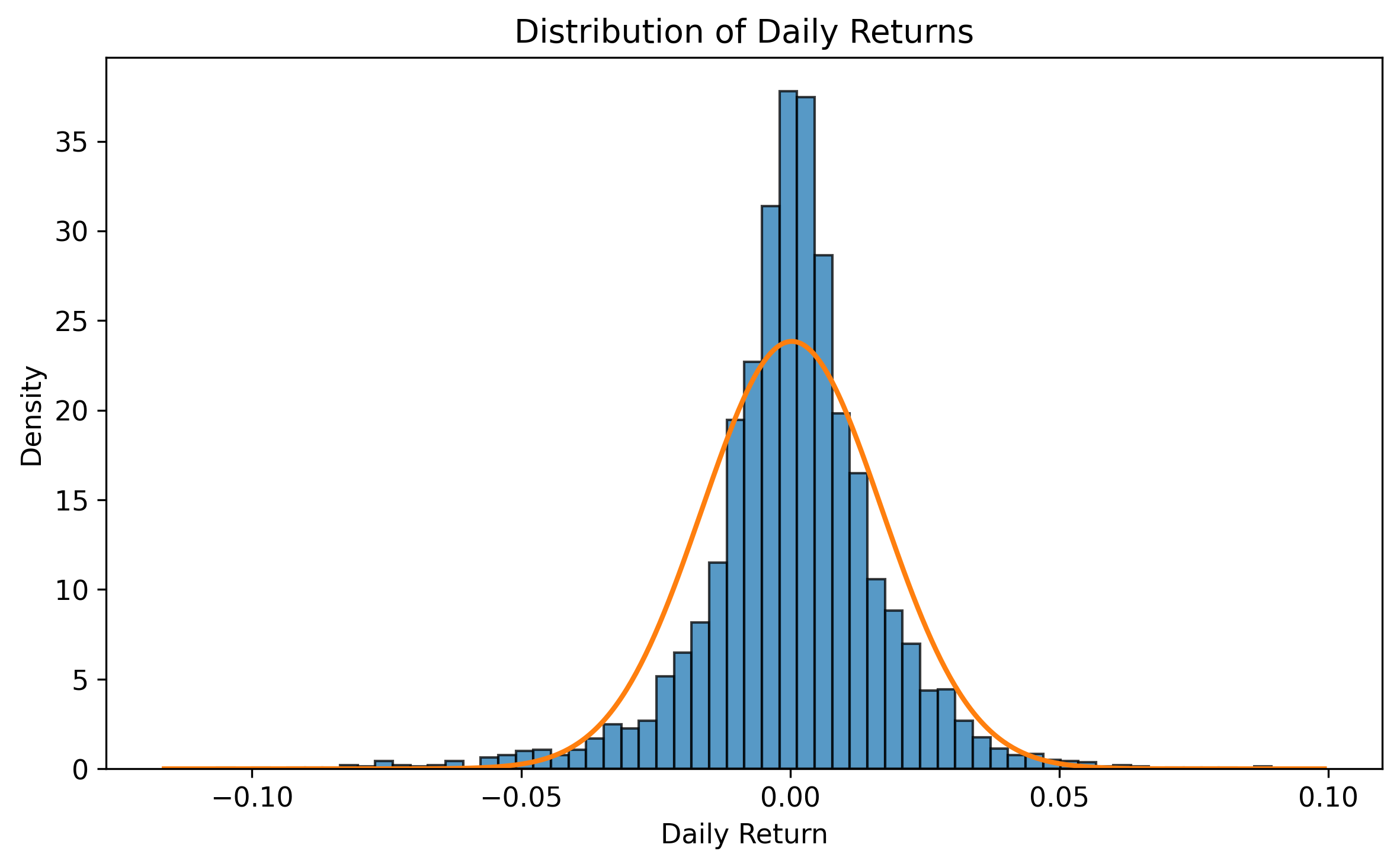}
\figurenote{The histogram shows the empirical distribution of daily returns together with a fitted normal density.}
\end{figure}

\begin{figure}[H]
\centering
\caption{QQ plot of daily returns}
\label{fig:return_qq}
\includegraphics[width=0.60\textwidth]{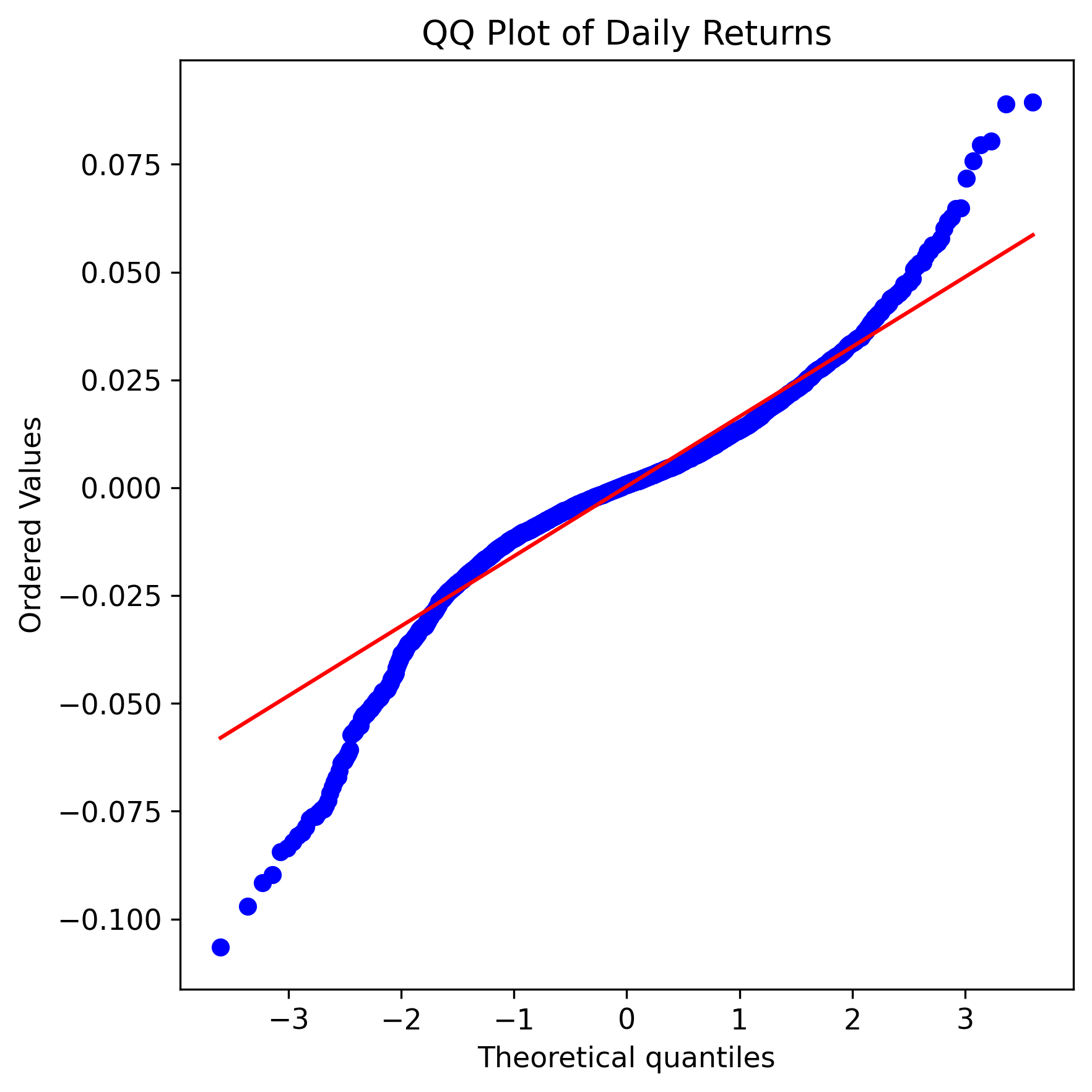}

\figurenote{The QQ plot compares empirical return quantiles with theoretical normal quantiles. Substantial tail deviations indicate strong non-normality.}
\end{figure}

Figures~\ref{fig:return_hist} and \ref{fig:return_qq} provide graphical evidence of these distributional characteristics. The histogram reveals a peaked distribution with heavy tails, while the QQ plot shows pronounced departures from the Gaussian benchmark, particularly in the extreme quantiles.

The distribution of realized variance is highly right-skewed and characterized by extreme kurtosis. After applying a logarithmic transformation, the distribution becomes substantially more symmetric and approximately Gaussian. Skewness declines markedly and kurtosis moves closer to the Gaussian benchmark of three.

\begin{figure}[H]
\centering
\caption{Distribution of log-realized variance}
\label{fig:logrv_hist}
\includegraphics[width=0.60\textwidth]{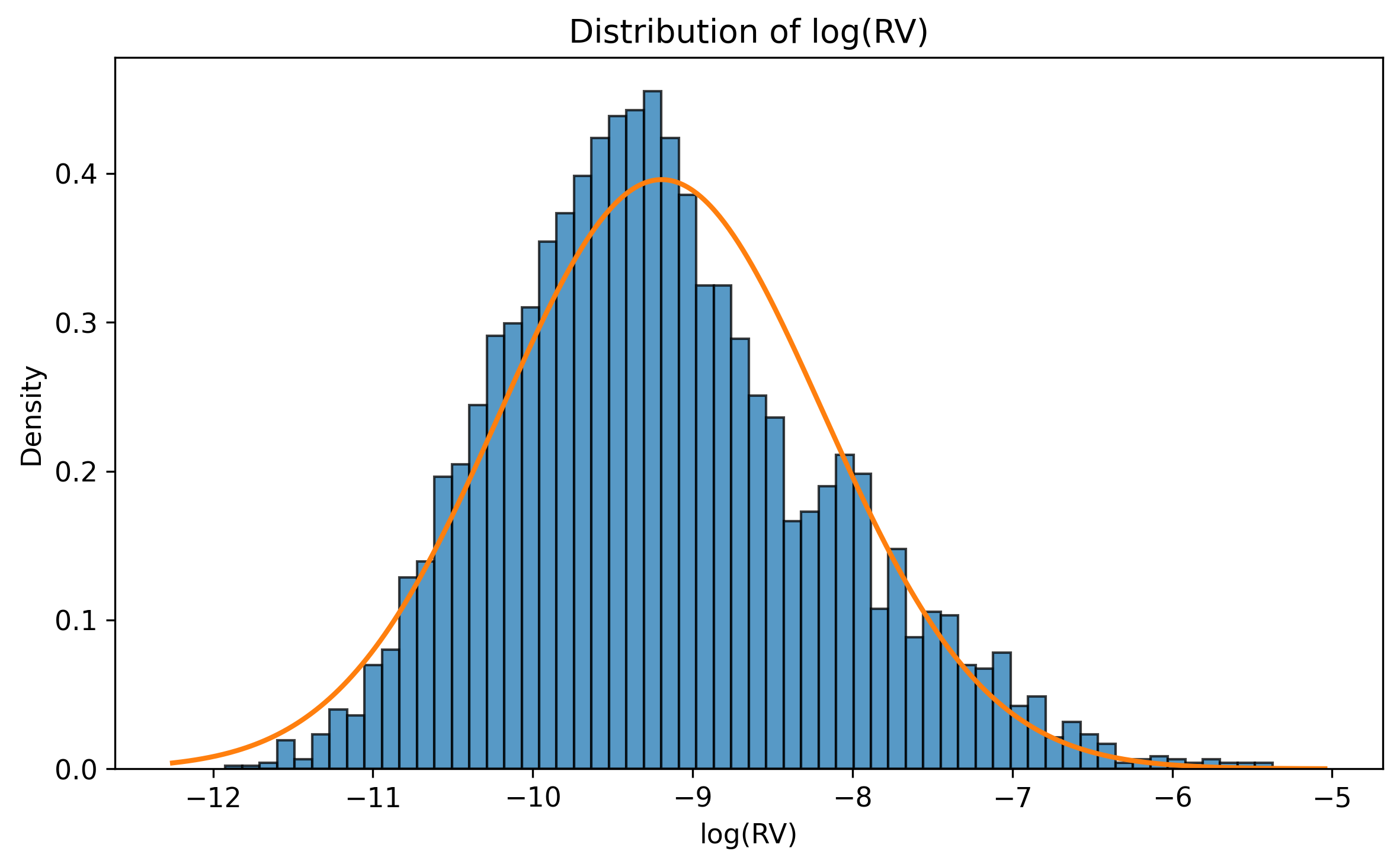}

\figurenote{The histogram shows the empirical distribution of log-realized variance together with a fitted normal density.}
\end{figure}

\begin{figure}[H]
\centering
\caption{QQ plot of log-realized variance}
\label{fig:logrv_qq}
\includegraphics[width=0.60\textwidth]{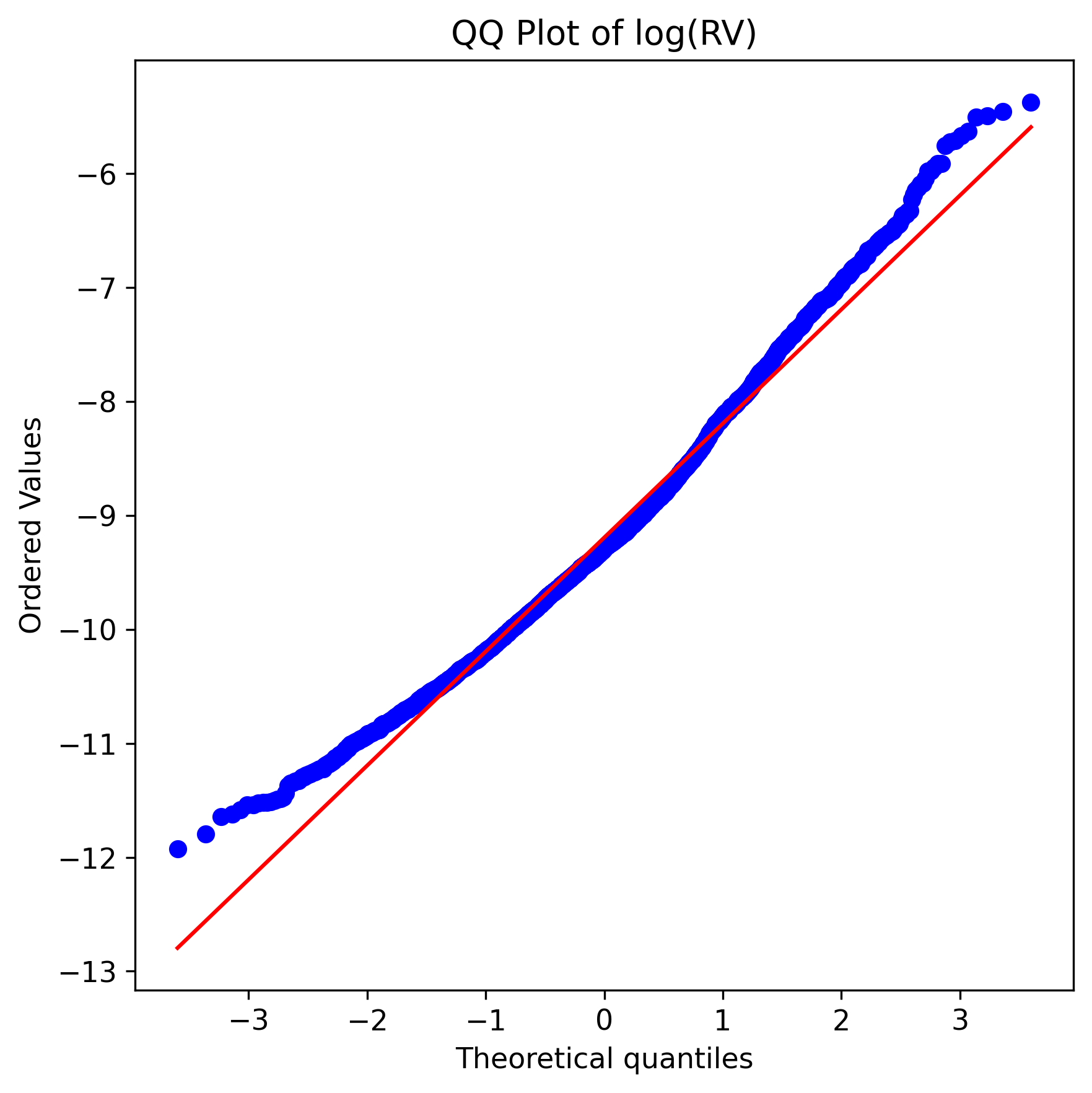}

\figurenote{The QQ plot compares empirical quantiles of log-realized variance with theoretical normal quantiles.}
\end{figure}

Figures~\ref{fig:logrv_hist} and \ref{fig:logrv_qq} illustrate the effectiveness of this transformation. Although some tail deviations remain, the log transformation considerably improves the statistical properties of realized volatility and facilitates subsequent linear modeling.

\begin{figure}[H]
\centering
\caption{Time series of log-realized variance}
\label{fig:logrv_ts}
\includegraphics[width=0.80\textwidth]{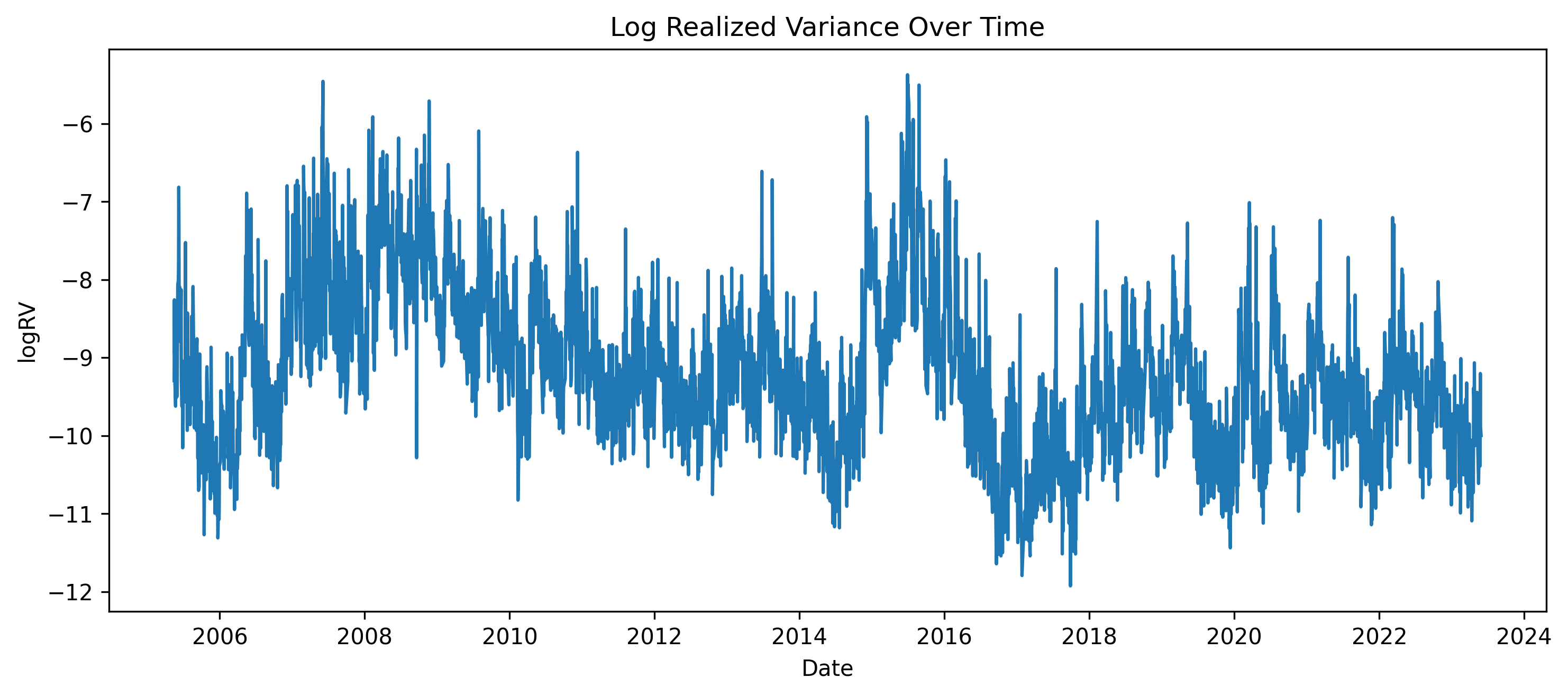}

\figurenote{The figure illustrates the time-series evolution of log-realized variance and highlights volatility clustering.}
\end{figure}

Beyond distributional characteristics, volatility exhibits strong temporal dependence. Figure~\ref{fig:logrv_ts} presents the time series of log-realized variance and reveals pronounced volatility clustering, whereby high-volatility periods tend to be followed by high-volatility periods and low-volatility periods persist over time.

\begin{figure}[H]
\centering
\caption{Autocorrelation function of log-realized variance}
\label{fig:logrv_acf}
\includegraphics[width=0.60\textwidth]{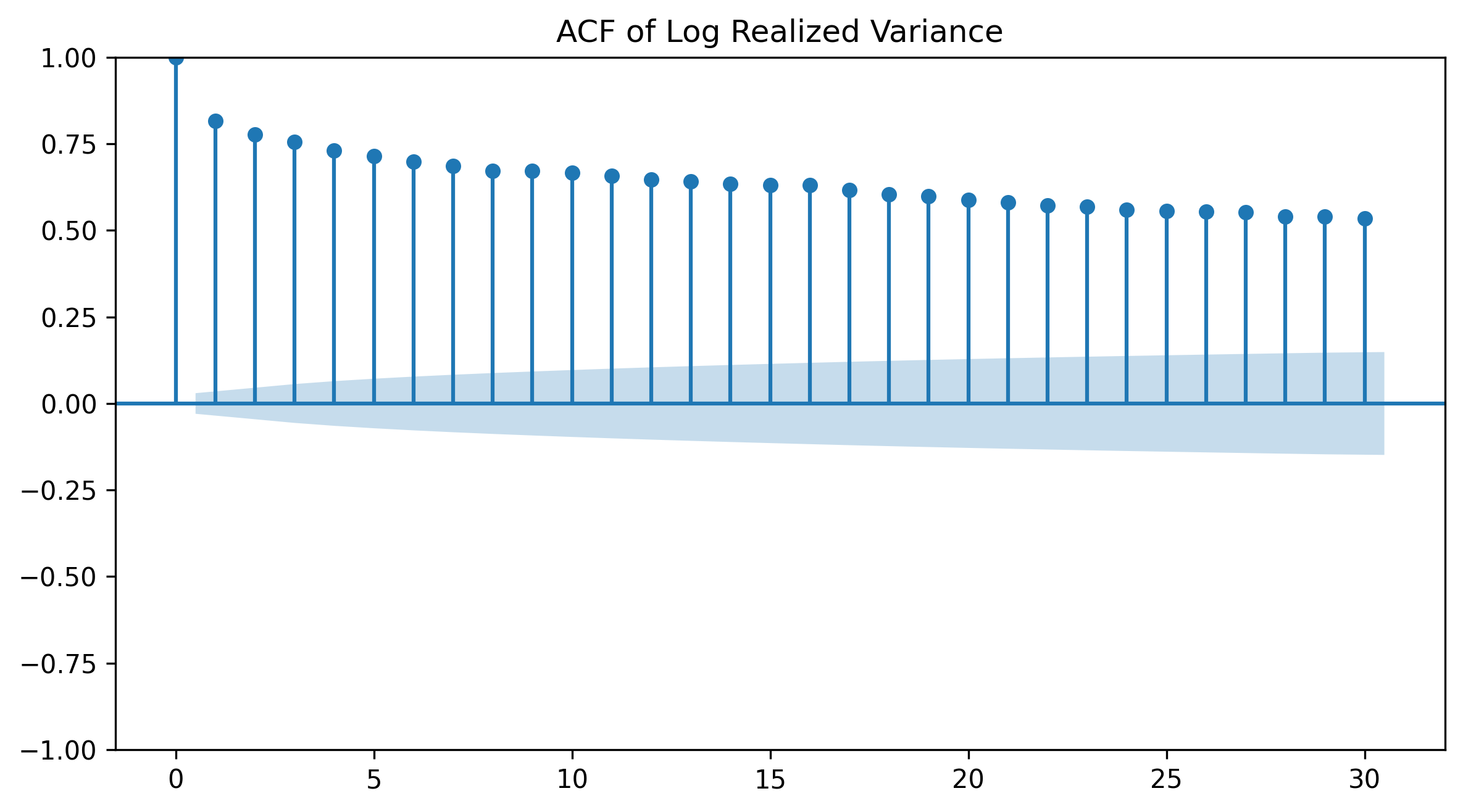}

\figurenote{The autocorrelation function displays strong persistence and slowly decaying dependence in log-realized variance.}
\end{figure}

Figure~\ref{fig:logrv_acf} reports the autocorrelation function of log-realized variance. Autocorrelations remain positive and decay slowly across lags, providing strong evidence of long-memory behavior in volatility dynamics.

Overall, the descriptive evidence highlights two key empirical features. First, returns are non-normal, asymmetric, and heavy-tailed. Second, realized volatility exhibits strong persistence, clustering, and long-memory behavior. These findings motivate the use of log-transformed realized volatility, HAR-type forecasting models, and heavy-tailed volatility specifications in the subsequent analysis.

\subsection{Walk-forward design}

To evaluate forecasting performance in a realistic investment setting, the study adopts an expanding-window walk-forward framework for both volatility forecasting and return prediction. This approach is closely related to recursive forecasting schemes commonly used in time-series analysis \citep{hamilton1994} and is widely regarded as the standard methodology for out-of-sample forecast evaluation in empirical finance \citep{timmermann2006,goyal2008,gu2020}.

At each re-estimation date, model parameters are estimated using all information available up to that point.Between re-estimation dates, model parameters remain fixed and forecasts are generated sequentially using newly available observations. One-step-ahead forecasts are subsequently generated for the next out-of-sample period. As new observations become available, the estimation window expands recursively and the forecasting procedure is repeated. This design mimics real-time forecasting and prevents the use of future information.

For volatility forecasting, the initial estimation window covers the first six years of observations. Under the baseline specification, model parameters are re-estimated every three months, and one-step-ahead volatility forecasts are generated sequentially throughout the out-of-sample period.

For return prediction, forecasting begins only after the minimum training requirement of $N_{\min}=300$ observations is satisfied. The initial estimation sample is divided into 180 training observations and 120 validation observations, allowing hyperparameter selection within each walk-forward iteration. Under the baseline specification, return-prediction models are re-estimated quarterly and generate strictly out-of-sample one-step-ahead forecasts.

\begin{figure}[H]
\centering
\caption{Expanding-window walk-forward design}
\label{fig:walkforward_design}
\includegraphics[width=0.95\textwidth]{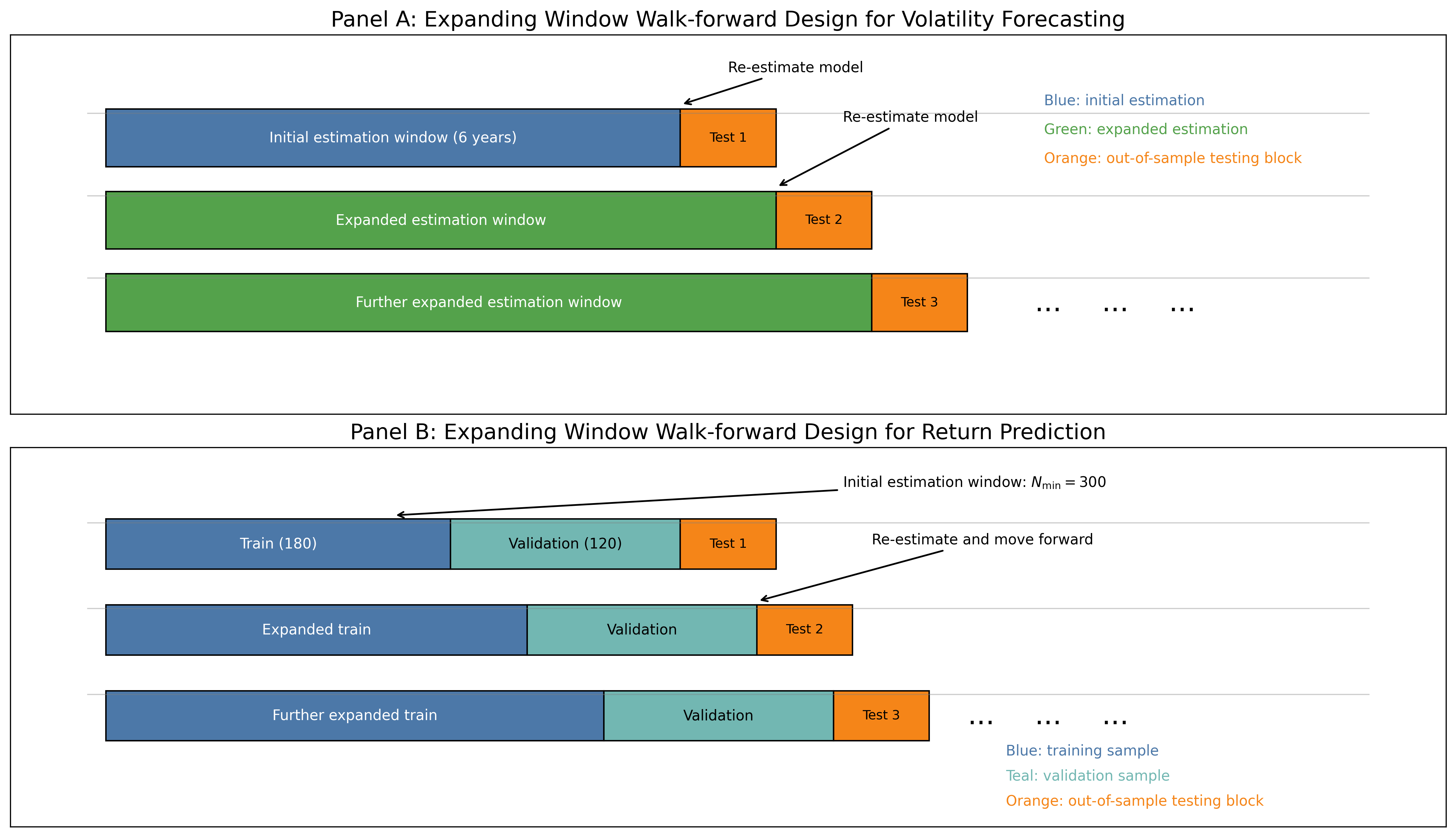}

\figurenote{Panel A illustrates the volatility-forecasting framework. Panel B illustrates the return-prediction framework. Models are re-estimated every three months under the baseline specification, and forecasts are generated strictly out of sample.}
\end{figure}

Figure~\ref{fig:walkforward_design} provides a schematic representation of the walk-forward framework. Panel A illustrates the volatility-forecasting design, while Panel B illustrates the return-prediction design. In both cases, the orange segments denote out-of-sample evaluation periods.

The effective out-of-sample periods differ across forecasting tasks because of data availability and model-estimation requirements.

For volatility forecasting, the availability of regime variables and the minimum estimation-sample requirement determine the start of the out-of-sample period. Consequently, the volatility-forecasting evaluation window extends from August 21, 2013 to May 29, 2023, yielding 2,326 out-of-sample observations.

For return prediction, all predictors, including volatility forecasts, regime variables, and lagged return features, must first be aligned into a common dataset. After merging these variables and applying the minimum training requirement, the feasible out-of-sample evaluation period runs from December 17, 2014 to May 26, 2023, yielding 2,004 observations.

The difference between the volatility-forecasting and return-prediction samples reflects feature-alignment constraints rather than differences in the underlying data source. Throughout the analysis, all forecasts are generated using information available at the time the forecast is made, ensuring a strictly out-of-sample and look-ahead-bias-free evaluation framework.

\section{Methodology}
\label{sec:methodology}

\subsection{Volatility forecasting framework}
\label{sec:vol_method}

The volatility forecasting framework is implemented as a sequential two-stage procedure that combines realized-volatility modeling with regime-switching conditional variance dynamics. The objective is to capture both the persistent component of volatility and potential structural changes in market conditions.

The framework consists of three steps. First, a HARQ model is estimated to capture the heterogeneous long-memory structure of realized volatility. Second, a Markov-switching GJR-GARCH model is fitted to the HARQ residuals in order to identify latent volatility regimes and extract filtered regime probabilities. Third, the resulting regime information is reintroduced into the HARQ forecasting equation to generate regime-augmented volatility forecasts.

Importantly, the procedure is sequential rather than jointly estimated. The HARQ model is estimated first, the residual process is modeled second, and the resulting regime probabilities are subsequently incorporated as exogenous forecasting inputs. Therefore, the framework should be interpreted as a forecasting architecture rather than a fully structural regime-switching system.

\subsubsection{HARQ model}

The conditional mean of log-realized variance is modeled using a heterogeneous autoregressive specification augmented with realized quarticity:
\begin{equation}
\log(RV_t)
=
\beta_0
+
\beta_1 \log(RV_{t-1})
+
\beta_2 \log(RV_{t-5:t-1})
+
\beta_3 \log(RV_{t-22:t-1})
+
\beta_4 \log(RQ_{t-1})
+
\varepsilon_t .
\label{eq:harq_main}
\end{equation}

Here, $RV_t$ denotes realized variance and $RQ_t$ denotes realized quarticity. The weekly and monthly components are defined as rolling averages of realized volatility over the previous five and twenty-two trading days, respectively. This specification captures the heterogeneous persistence structure of financial volatility and serves as the baseline volatility-forecasting model throughout the analysis.

\subsubsection{Residual-based regime identification}

To capture state-dependent volatility dynamics, the residuals from Equation~\eqref{eq:harq_main} are modeled using a two-state Markov-switching GJR-GARCH framework.

Rather than applying a regime-switching model directly to realized volatility, the analysis focuses on the residual component of the HARQ model. This design allows regime identification to capture state dependence beyond the persistence already explained by the HARQ specification.

The empirical implementation follows the MSGARCH framework of \citet{ardia2019msgarch}, which allows flexible estimation of regime-dependent GARCH models with asymmetric responses and heavy-tailed innovations.

Let $S_t \in \{1,2\}$ denote the latent volatility regime. Conditional on regime $S_t$, the residual variance evolves according to a GJR-GARCH process.

\subsubsection{Markov-switching GJR-GARCH model}

Conditional on the latent state $S_t$, the residual variance follows
\begin{equation}
\sigma_t^2
=
\omega_{S_t}
+
\alpha_{S_t}\varepsilon_{t-1}^{2}
+
\gamma_{S_t}
\mathbf{1}(\varepsilon_{t-1}<0)
\varepsilon_{t-1}^{2}
+
\beta_{S_t}\sigma_{t-1}^{2}.
\label{eq:msgjrgarch}
\end{equation}

The indicator term captures the leverage effect, allowing negative shocks to affect future volatility differently from positive shocks of the same magnitude.

To accommodate heavy tails, the standardized innovations are assumed to follow a Student-$t$ distribution:
\begin{equation}
\varepsilon_t = \sigma_t z_t,
\qquad
z_t \sim t_{\nu}.
\label{eq:student_t_innovations}
\end{equation}

This specification provides greater robustness to extreme observations than the Gaussian assumption.

\subsubsection{Filtered regime probabilities}

The latent-state probabilities are obtained using filtered rather than smoothed probabilities. Let
\begin{equation}
p_t(j)
=
P(S_t=j \mid \mathcal{F}_t)
\label{eq:filtered_prob}
\end{equation}
denote the filtered probability of state $j$ at time $t$, where $\mathcal{F}_t$ represents the information set available up to time $t$.

Filtered probabilities are used because they depend only on contemporaneously available information. By contrast, smoothed probabilities would incorporate future observations and therefore introduce look-ahead bias.

The high-volatility regime is identified using the probability-weighted average conditional volatility within each estimation window. Specifically,
\begin{equation}
j^*
=
\arg\max_j
\frac{\sum_t p_t(j)\sigma_t}
{\sum_t p_t(j)} .
\label{eq:highvol_state}
\end{equation}

The regime probability used throughout the subsequent analysis is therefore
\begin{equation}
p_t
=
P(S_t=j^* \mid \mathcal{F}_t),
\label{eq:highvol_prob}
\end{equation}
which represents the filtered probability that the market is in the high-volatility state.

\subsubsection{Regime-augmented HARQ forecasts}

The estimated regime probability is subsequently incorporated into the volatility forecasting equation as an exogenous predictor:
\begin{equation}
\begin{aligned}
\log(RV_t)
=&
\beta_0
+
\beta_1 \log(RV_{t-1})
+
\beta_2 \log(RV_{t-5:t-1})
\\
&
+
\beta_3 \log(RV_{t-22:t-1})
+
\beta_4 \log(RQ_{t-1})
+
\gamma p_t
+
\varepsilon_t .
\end{aligned}
\label{eq:harq_pt}
\end{equation}

Unlike discrete regime classification, the continuous probability $p_t$ allows gradual transitions between market states and avoids abrupt regime switching.

The resulting volatility forecasts are used as first-stage outputs and subsequently enter the machine-learning return-prediction model.

\subsection{Return prediction framework}
\label{sec:return_method}

The second stage of the empirical framework focuses on forecasting next-day returns using machine learning. While volatility forecasting has traditionally achieved relatively strong predictive performance, return prediction remains substantially more challenging because of the low signal-to-noise ratio inherent in financial returns. To accommodate nonlinear relationships and complex interactions among predictors, this study employs Extreme Gradient Boosting (XGBoost), a tree-based ensemble learning algorithm that has demonstrated strong predictive performance in financial applications.

The objective is to estimate the conditional expectation of future returns using information available at time $t$:
\begin{equation}
\mathbb{E}(r_{t+1}\mid X_t)=f(X_t),
\label{eq:return_conditional_expectation}
\end{equation}
where $r_{t+1}$ denotes the next-day return, $X_t$ represents the predictor set observed at time $t$, and $f(\cdot)$ is an unknown nonlinear function estimated by XGBoost.

The resulting forecast is
\begin{equation}
\hat r_{t+1}=\hat f(X_t).
\label{eq:return_forecast}
\end{equation}

All forecasts are generated strictly out of sample. Consequently, the prediction target is always $r_{t+1}$, whereas all predictors are constructed using information available no later than time $t$.

\subsubsection{XGBoost model}

XGBoost \citep{chen2016} is an ensemble-learning algorithm based on gradient-boosted decision trees. Rather than relying on a single predictive model, XGBoost constructs an additive ensemble of regression trees:
\begin{equation}
\hat r_{t+1}
=
\sum_{k=1}^{K}
f_k(X_t),
\label{eq:xgb_ensemble}
\end{equation}
where each $f_k(\cdot)$ represents an individual regression tree and $K$ denotes the number of trees.

Model estimation is obtained by minimizing a regularized objective function:
\begin{equation}
\mathcal{L}
=
\sum_{i=1}^{N}
l(r_i,\hat r_i)
+
\sum_{k=1}^{K}
\Omega(f_k),
\label{eq:xgb_objective}
\end{equation}
where $l(\cdot)$ denotes the prediction loss and $\Omega(\cdot)$ is a regularization term controlling model complexity.

Compared with traditional linear forecasting models, XGBoost can accommodate nonlinear relationships, threshold effects, and interactions among predictors without requiring explicit model specification. These properties make it suitable for financial prediction problems where predictive relationships are often unstable and state dependent.

\subsubsection{Predictor construction}

The predictor set combines volatility information, regime information, and lagged return dynamics. The final specification contains twelve predictors:
\begin{equation}
\begin{aligned}
X_t = (&
\texttt{logRVhat\_t1},
\texttt{vol\_lag5},
\texttt{vol\_lag22},
\log(RV_t),
\log(RQ_t),
\texttt{signed\_jump}_t,
\\
&
p_t,
\texttt{pt\_lag5},
\texttt{vol\_pt},
\texttt{ret\_lag5},
\texttt{ret\_lag22},
\texttt{ret\_abs\_lag1}
).
\end{aligned}
\label{eq:predictor_set}
\end{equation}

The first group consists of volatility-related variables. These variables capture both current and expected market risk conditions. Among them, the forecasted volatility measure \texttt{logRVhat\_t1} is generated by the first-stage volatility forecasting framework. Consequently, the return-prediction model explicitly incorporates information extracted from the regime-dependent HARQ--MS--GJR--GARCH--$t$ volatility model.

Because \texttt{logRVhat\_t1} is itself an estimated quantity rather than a directly observed variable, it should be interpreted as a generated regressor. As a result, its predictive contribution depends on the quality of the preceding volatility-forecasting stage and may contain estimation error inherited from first-stage model estimation.

The second group consists of regime-related variables. The filtered probability $p_t$ measures the probability that the market is currently in the high-volatility regime identified by the Markov-switching model. The lagged probability \texttt{pt\_lag5} captures short-run persistence in regime dynamics, while the interaction term \texttt{vol\_pt} allows the predictive role of volatility to vary across market states.

Importantly, only filtered probabilities are used throughout the analysis. Smoothed probabilities are excluded because they condition on future observations and would therefore introduce look-ahead bias.

The third group contains lagged-return variables. These predictors capture short-horizon momentum, reversal effects, and recent market activity. Together, the three groups of variables provide a flexible representation of volatility conditions, regime dependence, and return dynamics.

\subsubsection{Walk-forward estimation and model selection}

Model estimation follows the expanding-window walk-forward framework described in Section~\ref{sec:data}. Forecasting begins only after the minimum training requirement of $N_{\min}=300$ observations is satisfied.

At each re-estimation date, all observations available up to time $t$ are used for model estimation. The available sample is divided into a training subset and a validation subset. Under the baseline specification, the initial estimation window contains 180 training observations and 120 validation observations.

Hyperparameters are selected using validation performance rather than in-sample fit. The objective is to maximize predictive correlation on the validation sample, thereby emphasizing economically relevant forecasting ability rather than minimizing prediction error alone.

To reduce overfitting and maintain computational tractability, a deliberately conservative hyperparameter grid is employed. The candidate specifications vary primarily along three dimensions: tree depth, learning rate, and regularization strength. This restricted search design reflects the weak-predictability nature of financial returns and reduces the likelihood of selecting hyperparameters that merely fit validation noise.

The baseline return-prediction framework uses quarterly (3M) model re-estimation. Alternative re-estimation frequencies of one month and six months are considered separately as robustness checks. The quarterly specification is retained as the main baseline specification because it balances statistical predictability, economic performance, and parameter stability across the out-of-sample period.

Although XGBoost is a nonlinear machine-learning model, feature-importance measures are reported to assess the relative contribution of different predictor groups. In addition, feature-ablation experiments are conducted to evaluate the incremental value of volatility forecasts and regime-related variables for both statistical return prediction and economic strategy performance.

The resulting return forecasts constitute the primary input for the trading strategies examined below.

\subsubsection{Hyperparameter selection}

Hyperparameters for the XGBoost return-prediction model are selected within each walk-forward estimation window using validation performance. The available estimation sample is split into a training subset and a validation subset, and the candidate specification that maximizes validation predictive correlation is selected.

To reduce overfitting in a weak-signal environment, the hyperparameter search is deliberately restricted to a small and economically interpretable grid. The grid varies model complexity, learning speed, and regularization strength through tree depth, learning rate, $\gamma$, and $\lambda$. This conservative design limits the risk that the model selects configurations that fit validation noise rather than persistent predictive structure.

The baseline grid is defined as:
\begin{equation}
\begin{aligned}
\mathcal{G}
=
\{
&(300,2,0.05,3,0.0,1.0),\\
&(300,3,0.05,3,0.0,1.0),\\
&(300,2,0.03,3,0.0,1.0),\\
&(300,2,0.05,3,0.1,2.0)
\}.
\end{aligned}
\label{eq:xgb_grid}
\end{equation}
Each tuple denotes
\[
(n_{\text{estimators}},
\text{max\_depth},
\text{learning\_rate},
\text{min\_child\_weight},
\gamma,
\lambda).
\]

If all candidate models yield negative validation correlation, the procedure defaults to the most conservative specification with stronger regularization, and the validation correlation is treated as zero. This prevents the selection of models that display unstable or spurious negative predictive relationships.

An expanded hyperparameter grid is examined as a robustness exercise, but it does not improve out-of-sample performance. Therefore, the restricted grid is retained as the main specification. Detailed hyperparameter-selection frequencies are reported in the appendix rather than in the main results section.

\subsection{Trading strategy design}
\label{sec:strategy}

The primary objective of the trading-strategy analysis is to evaluate whether statistically significant predictive information can be transformed into economically meaningful investment performance. Rather than focusing solely on forecast accuracy, the analysis explicitly incorporates portfolio construction, transaction costs, turnover control, and regime dependence.

The main implementation considered in this study is a Low-vol Gated Weekly Signal$\times$Risk strategy. The design is motivated by three empirical observations. First, return predictability is generally weak. Second, predictive performance is concentrated primarily in low-volatility market states. Third, naive trading strategies tend to lose economic value once transaction costs are considered. Consequently, the strategy is designed as a conservative, low-turnover implementation that emphasizes risk control and selective participation rather than aggressive return maximization.

\subsubsection{Risk-adjusted signal construction}

Let $\hat r_{t+1}$ denote the one-step-ahead return forecast generated by the XGBoost model and let $\widehat{\log RV}_{t+1}$ denote the forecasted log-realized variance from the volatility model.

Forecasted volatility is defined as
\begin{equation}
\hat \sigma_{t+1}
=
\sqrt{\exp(\widehat{\log RV}_{t+1})}.
\label{eq:forecast_vol}
\end{equation}

The raw trading signal is then constructed as
\begin{equation}
s_t
=
\frac{\hat r_{t+1}}
{\hat \sigma_{t+1}}.
\label{eq:signalrisk}
\end{equation}

This specification scales expected return by forecasted risk and therefore increases exposure when return forecasts are favorable relative to anticipated volatility.

\subsubsection{Low-volatility regime gating}

The empirical analysis shows that predictive performance is concentrated primarily in low-volatility market states. To exploit this state dependence, exposure is conditioned on the filtered regime probability obtained from the Markov-switching volatility model.

Let $p_t$ denote the filtered probability of the high-volatility regime. The gated signal is defined as
\begin{equation}
s_t^{G}
=
\begin{cases}
s_t,
&
p_t \leq 0.5,
\\
\kappa s_t,
&
p_t > 0.5,
\end{cases}
\label{eq:gated_signal}
\end{equation}
where $\kappa \in [0,1]$ controls exposure during high-volatility periods.

This design reflects the empirical finding that high-volatility regimes contain limited directional predictability and are therefore treated primarily as exposure-suppression states rather than alpha-generation opportunities.

\subsubsection{Signal filtering}

Because return forecasts are noisy, a thresholding procedure is applied to eliminate weak signals. Let
\begin{equation}
Q_{q,t}^{WF}
\left(
|s^{G}|
\right)
\label{eq:wf_quantile}
\end{equation}
denote the walk-forward estimate of the $q$-th quantile of the absolute gated signal, computed using only information available before time $t$.

The filtered signal is then
\begin{equation}
\tilde s_t
=
\begin{cases}
s_t^{G},
&
|s_t^{G}|
\ge
Q_{q,t}^{WF},
\\
0,
&
\text{otherwise}.
\end{cases}
\label{eq:filtered_signal}
\end{equation}

Under the baseline specification, the threshold parameter is set to
\begin{equation}
q=0.60.
\label{eq:q_baseline}
\end{equation}

Consequently, only the strongest 40\% of signals are retained, while weaker signals are discarded. Importantly, threshold calibration is performed recursively using only historical information. This ensures that no future information enters the signal-selection process.

\subsubsection{Walk-forward position scaling}

After thresholding, the remaining signals are scaled to achieve stable portfolio exposure through a walk-forward calibration procedure.

At each date $t$, a scaling coefficient $c_t^{WF}$ is estimated using historical filtered signals available prior to time $t$. The objective is to target an average absolute exposure of approximately 0.5:
\begin{equation}
\frac{1}{T_t}
\sum_{\tau<t}
|c_t^{WF}\tilde s_\tau|
\approx 0.5.
\label{eq:scaling_objective}
\end{equation}

Target portfolio weights are then defined as
\begin{equation}
w_t^{*}
=
\min
\Bigl(
\max(c_t^{WF}\tilde s_t,-w_{\max}),
w_{\max}
\Bigr),
\label{eq:target_weight}
\end{equation}
where
\begin{equation}
w_{\max}=0.6.
\label{eq:wmax}
\end{equation}

The position cap prevents extreme exposures and improves implementation stability. Because thresholding sets many signals to zero and the cap truncates extreme positions, realized average exposure may be lower than the nominal target. The exposure target should therefore be interpreted as a calibration objective rather than a guaranteed portfolio characteristic.

\subsubsection{Weekly rebalancing and turnover control}

The strategy is implemented using weekly rebalancing. Between rebalancing dates, existing positions are carried forward.

To further reduce unnecessary trading activity, a no-trade band is imposed. Let $w_{t-1}$ denote the previous portfolio weight and $w_t^{*}$ denote the target weight. The implemented portfolio weight is
\begin{equation}
w_t =
\begin{cases}
w_{t-1},
&
\text{if } |w_t^{*}-w_{t-1}|<b,
\\
w_t^{*},
&
\text{otherwise},
\end{cases}
\label{eq:implemented_weight}
\end{equation}
where the baseline specification uses
\begin{equation}
b=0.02.
\label{eq:no_trade_band}
\end{equation}

This mechanism suppresses minor portfolio adjustments and substantially reduces turnover.

\subsubsection{Portfolio returns and transaction costs}

Portfolio turnover is defined as
\begin{equation}
\mathrm{Turnover}_t =
|w_t-w_{t-1}|.
\label{eq:turnover}
\end{equation}

Gross portfolio returns are computed as
\begin{equation}
R_t^{\mathrm{gross}} =
w_t r_{t+1},
\label{eq:gross_return}
\end{equation}
where $r_{t+1}$ denotes the realized next-period return.

Net returns incorporate transaction costs:
\begin{equation}
R_t^{\mathrm{net}}
=
R_t^{\mathrm{gross}}
-
c_{\mathrm{cost}}
\cdot
\mathrm{Turnover}_t .
\label{eq:net_return}
\end{equation}

The baseline specification assumes a one-way transaction cost of
\begin{equation}
c_{\mathrm{cost}} =
0.0005,
\label{eq:cost_baseline}
\end{equation}
corresponding to 5 basis points. Additional transaction-cost levels are examined in the robustness analysis to evaluate implementation sensitivity.

\subsubsection{Baseline specification}

Unless otherwise stated, the baseline strategy employs:
\begin{itemize}
\item quarterly (3M) model re-estimation,
\item weekly portfolio rebalancing,
\item threshold quantile $q=0.60$,
\item no-trade band $b=0.02$,
\item target mean absolute exposure of 0.5,
\item maximum portfolio weight $w_{\max}=0.6$,
\item one-way transaction costs of 5 basis points.
\end{itemize}

These parameter values are treated as conservative ex-ante choices rather than ex-post optimized values. Alternative parameter combinations are examined separately through sensitivity analyses and multiple-testing robustness procedures.

\subsubsection{Strategy parameter selection}

The baseline strategy parameters are selected ex ante rather than through ex post performance optimization. The threshold quantile ($q=0.60$), no-trade band ($b=0.02$), target mean absolute exposure, maximum position cap, weekly rebalancing frequency, and 5 bp transaction-cost assumption are treated as conservative implementation choices designed to balance signal strength, turnover control, and economic interpretability.

The threshold parameter $q=0.60$ retains only relatively strong signals while avoiding excessive sparsity in portfolio exposure. The no-trade band $b=0.02$ suppresses small portfolio adjustments and reduces unnecessary turnover. The target mean absolute exposure of 0.5 and position cap of $w_{\max}=0.6$ are used to prevent excessive leverage and improve implementation stability.

These choices are not interpreted as statistically optimal parameters. Alternative parameter configurations are evaluated separately through sensitivity analyses over the $q \times b$ grid and through multiple-testing procedures. This distinction is important because the main strategy is intended to represent a conservative baseline implementation rather than the best-performing ex post configuration.

\subsubsection{Benchmark and alternative implementations}

To evaluate whether the proposed strategy provides incremental economic value, several benchmark and alternative implementations are also considered.

The benchmark set includes:
\begin{itemize}
\item Buy-and-Hold,
\item Volatility-managed Buy-and-Hold,
\item Momentum Weekly,
\item Regime-only Long-Flat,
\item Regime-only Long-Short.
\end{itemize}

These benchmarks represent increasingly sophisticated alternatives based on passive exposure, volatility scaling, trend-following, and regime timing.

In addition, several signal-decomposition strategies are evaluated to better understand the sources of economic value embedded in the proposed framework. These include:
\begin{itemize}
\item Return-only,
\item Volatility-only,
\item Signal$\times$Risk (ungated),
\item Long-only implementations.
\end{itemize}

The decomposition analysis allows the contribution of return forecasts, volatility forecasts, and regime information to be examined separately. Results for these benchmark and alternative implementations are reported in the robustness section.

\subsection{Evaluation metrics}
\label{sec:evaluation}

The empirical framework is evaluated from three complementary perspectives: volatility-forecasting performance, return-prediction performance, and economic strategy performance. This multi-layered evaluation design allows statistical forecasting accuracy and economic value to be assessed jointly.

\subsubsection{Volatility-forecasting evaluation}

The out-of-sample performance of volatility forecasts is evaluated using three standard loss functions: mean squared error (MSE), quasi-likelihood loss (QLIKE), and log mean absolute error (LMAE).

Mean squared error in log space is defined as
\begin{equation}
\mathrm{MSE}_{\log}
=
\frac{1}{T}
\sum_{t=1}^{T}
\left[
\widehat{\log RV}_{t}
-
\log(RV_t)
\right]^2 .
\label{eq:mse_log}
\end{equation}

Log mean absolute error is given by
\begin{equation}
\mathrm{LMAE}
=
\frac{1}{T}
\sum_{t=1}^{T}
\left|
\widehat{\log RV}_{t}
-
\log(RV_t)
\right| .
\label{eq:lmae}
\end{equation}

To complement these symmetric loss functions, quasi-likelihood loss is computed as
\begin{equation}
\mathrm{QLIKE}
=
\frac{1}{T}
\sum_{t=1}^{T}
\left[
\log(\widehat{RV}_t)
+
\frac{RV_t}
{\widehat{RV}_t}
\right].
\label{eq:qlike}
\end{equation}

QLIKE is particularly attractive in volatility forecasting because it remains robust when realized volatility is an imperfect proxy for latent volatility and penalizes volatility underprediction relatively strongly.

Beyond forecast losses, forecast calibration is assessed using Mincer--Zarnowitz (MZ) regressions:
\begin{equation}
\log(RV_t)
=
\alpha
+
\beta \widehat{\log RV}_{t}
+
u_t .
\label{eq:mz_regression}
\end{equation}

Under perfect forecast calibration, the joint null hypothesis is
\begin{equation}
H_0:
\alpha = 0,
\qquad
\beta = 1 .
\label{eq:mz_null}
\end{equation}

Forecast superiority across competing volatility models is evaluated using Diebold--Mariano (DM) tests based on QLIKE loss differentials \citep{diebold1995}. Heteroskedasticity- and autocorrelation-consistent (HAC) standard errors are computed using the Newey--West estimator \citep{newey1987}. In addition to full-sample comparisons, regime-conditional DM tests are conducted to investigate whether regime-aware volatility models deliver greater forecasting improvements during high-volatility periods.

\subsubsection{Return-prediction evaluation}

Return-prediction performance is evaluated using measures of forecast accuracy, directional predictability, ranking ability, and temporal stability.

The primary measure of predictive accuracy is the Pearson correlation between predicted and realized returns:
\begin{equation}
\rho
=
\mathrm{corr}
\left(
\hat r_{t+1},
r_{t+1}
\right).
\label{eq:pearson_corr}
\end{equation}

Forecast magnitude errors are assessed using mean absolute error:
\begin{equation}
\mathrm{MAE}
=
\frac{1}{T}
\sum_{t=1}^{T}
\left|
\hat r_{t+1}
-
r_{t+1}
\right|.
\label{eq:return_mae}
\end{equation}

Directional predictability is evaluated using the hit ratio:
\begin{equation}
\mathrm{Hit}
=
\frac{1}{T}
\sum_{t=1}^{T}
\mathbf{1}
\left(
\hat r_{t+1}
\cdot
r_{t+1}
>
0
\right).
\label{eq:hit_ratio}
\end{equation}

Statistical significance of directional accuracy is evaluated using HAC-corrected inference rather than simple binomial tests, thereby accommodating serial dependence and conditional heteroskedasticity in forecast errors.

To evaluate ranking ability, Spearman rank correlation is also reported:
\begin{equation}
\rho^{S}
=
\mathrm{corr}
\left(
\mathrm{rank}(\hat r_{t+1}),
\mathrm{rank}(r_{t+1})
\right).
\label{eq:spearman_corr}
\end{equation}

Because financial predictability is often unstable over time, rolling Information Coefficients (ICs) are additionally computed:
\begin{equation}
IC_t^{(W)}
=
\mathrm{corr}
\left(
\{\hat r_{\tau+1}\}_{\tau=t-W+1}^{t},
\{r_{\tau+1}\}_{\tau=t-W+1}^{t}
\right),
\label{eq:rolling_ic}
\end{equation}
where $W$ denotes the rolling-window length.

The rolling IC analysis is summarized using:
\begin{itemize}
\item mean rolling IC,
\item rolling IC standard deviation,
\item information coefficient information ratio (ICIR),
\item proportion of positive-IC windows,
\item proportion of windows with economically meaningful positive IC.
\end{itemize}

Finally, all return-prediction metrics are reported both for the full sample and conditionally on volatility regimes. Specifically, predictive performance is examined separately during low-volatility states ($p_t \leq 0.5$) and high-volatility states ($p_t > 0.5$) to evaluate state-dependent predictability.

\subsubsection{Economic strategy evaluation}

The economic value of forecasts is evaluated through out-of-sample trading strategies.

Portfolio performance is assessed using annualized return compounded (ARC):
\begin{equation}
\mathrm{ARC}
=
\left(
\prod_{t=1}^{T}
(1+R_t)
\right)^{252/T}
-1 .
\label{eq:arc}
\end{equation}

Annualized standard deviation (ASD) is defined as
\begin{equation}
\mathrm{ASD}
=
\sqrt{252}
\times
\mathrm{Std}(R_t).
\label{eq:asd}
\end{equation}

Risk-adjusted performance is measured using the Sharpe ratio:
\begin{equation}
\mathrm{Sharpe}
=
\frac{\mathrm{ARC}}
{\mathrm{ASD}},
\label{eq:sharpe}
\end{equation}
and the Sortino ratio:
\begin{equation}
\mathrm{Sortino}
=
\frac{\mathrm{ARC}}
{\mathrm{DownsideDev}^{ann}},
\label{eq:sortino}
\end{equation}
where $\mathrm{DownsideDev}^{ann}$ denotes annualized downside deviation.

Downside risk is assessed using maximum drawdown:
\begin{equation}
\mathrm{MaxDD}
=
\min_t
\left(
\frac{W_t}
{\max_{s\leq t}W_s}
-1
\right),
\label{eq:maxdd}
\end{equation}
where cumulative wealth is defined as
\begin{equation}
W_t
=
\prod_{i=1}^{t}
(1+R_i).
\label{eq:wealth}
\end{equation}

Maximum loss duration (MLD) is also reported to measure the longest period during which portfolio value remains below its previous peak.

Benchmark-relative performance is evaluated using active-return information ratios:
\begin{equation}
IR^{\ast\ast}
=
\frac{
252 \cdot
\overline{(R_t-R_t^{BH})}
}
{
\sqrt{252}
\cdot
\mathrm{Std}(R_t-R_t^{BH})
},
\label{eq:ir_active}
\end{equation}
and downside-risk-adjusted information ratios:
\begin{equation}
IR^{\ast\ast\ast}
=
\frac{
252 \cdot
\overline{(R_t-R_t^{BH})}
}
{
DTE^{ann}
},
\label{eq:ir_downside}
\end{equation}
where $DTE^{ann}$ denotes annualized downside tracking error relative to the Buy-and-Hold benchmark.

Trading activity is evaluated using average turnover and the total number of trades. In addition, break-even transaction costs are reported. The break-even cost represents the transaction-cost level at which the strategy's Sharpe ratio falls to zero.

To assess statistical significance, circular block bootstrap procedures \citep{ledoit2008} are employed. First, each strategy's Sharpe ratio is tested against the null hypothesis of zero performance. Second, Sharpe-difference tests compare the proposed strategy against benchmark alternatives.

Finally, because multiple strategy configurations are considered throughout the robustness analysis, data-snooping concerns are addressed using White Reality Check-style and Hansen SPA-style procedures. These tests evaluate whether the apparent superiority of the best-performing strategy could arise from repeated model or parameter selection.

Taken together, this evaluation framework allows statistical forecasting performance, economic value, implementation robustness, and multiple-testing considerations to be assessed within a unified empirical setting.

\section{Empirical results}
\label{sec:results}

\subsection{Volatility forecasting results}
\label{sec:vol_results}

This subsection evaluates the out-of-sample performance of the regime-augmented HARQ volatility-forecasting framework. The main volatility-forecasting evaluation window spans from 2013-08-21 to 2023-05-29 and contains 2,326 observations. All models are evaluated on the same aligned out-of-sample window to ensure comparability.

\subsubsection{Regime identification}

\begin{figure}[H]
\centering
\caption{Data-driven identification of high-volatility regimes}
\label{fig:regime_identification}
\includegraphics[width=0.90\textwidth]{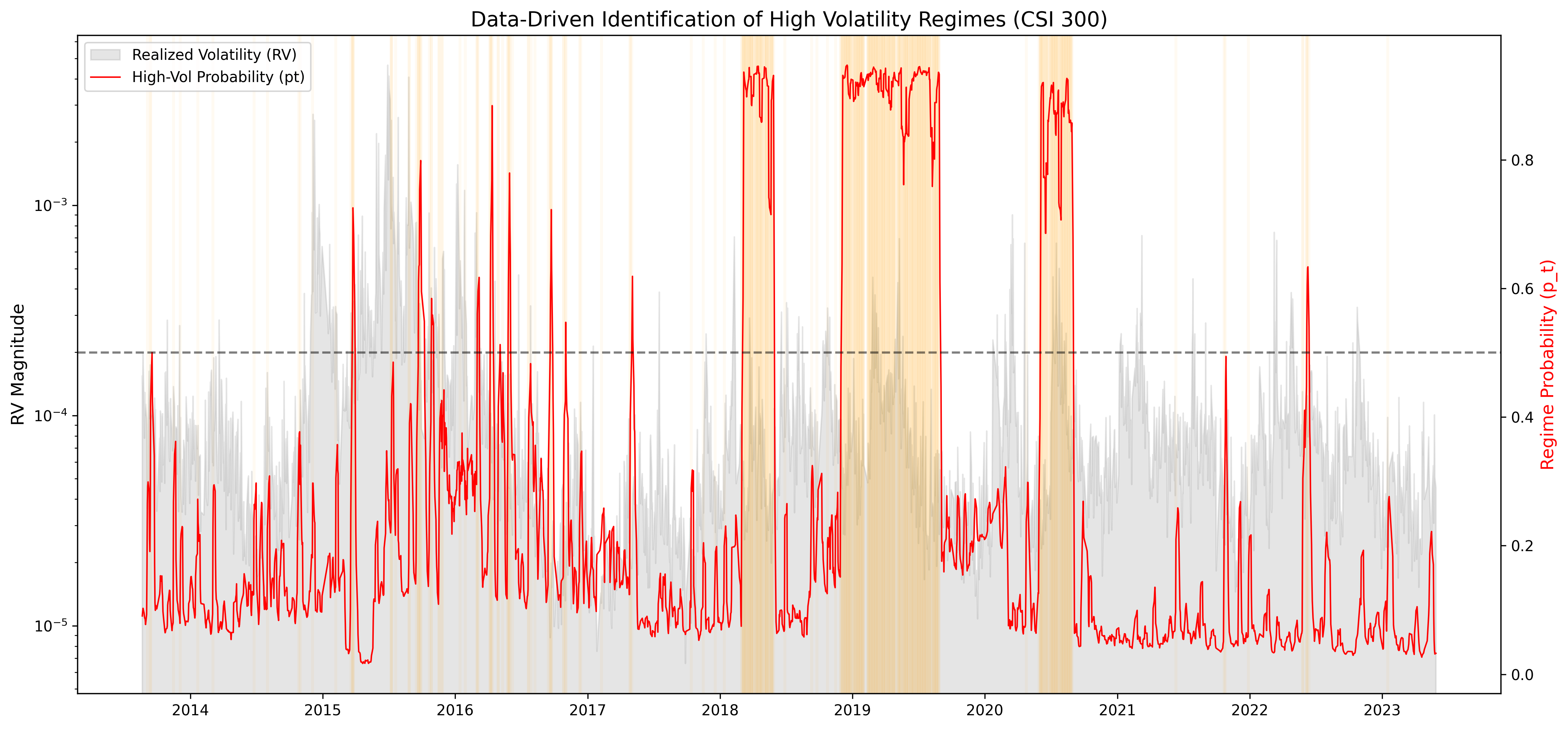}

\figurenote{The grey area represents realized volatility, while the red line denotes the filtered probability of the high-volatility regime ($p_t$). Orange shaded regions indicate periods where $p_t>0.5$, corresponding to identified high-volatility states. Filtered rather than smoothed probabilities are used to avoid look-ahead bias.}
\end{figure}

Figure~\ref{fig:regime_identification} presents the time-series dynamics of realized volatility together with the filtered probability of the high-volatility regime, denoted by $p_t$.

The estimated regime probability exhibits clear clustering behavior. Periods of elevated $p_t$ coincide with well-known episodes of market stress in the Chinese equity market, including the 2015--2016 stock market crash, the 2018--2019 trade-war period, and the COVID-related volatility episode in 2020. These periods are characterized not only by isolated volatility spikes, but also by persistent volatility clustering.

This distinction is important. Realized volatility measures the magnitude of current market fluctuations, whereas the regime probability captures the persistence and state dependence of volatility conditions. The filtered regime probability therefore provides a smoother and more economically interpretable measure of market stress, which motivates its use in the volatility-forecasting model and in the subsequent return-prediction and trading-strategy analysis.

\subsubsection{Out-of-sample forecasting performance}

Table~\ref{tab:vol_forecast_main} reports the out-of-sample volatility-forecasting performance of the baseline HARQ model and the regime-augmented HARQ + $p_t$ specification. Forecast accuracy is evaluated using mean squared error in log space, QLIKE loss, and log mean absolute error.

\begin{table}[H]
\centering
\footnotesize
\setlength{\tabcolsep}{5pt}
\renewcommand{\arraystretch}{1.05}
\caption{Out-of-sample volatility forecasting performance}
\label{tab:vol_forecast_main}

\begin{tabular}{lccc}
\toprule
Model & MSE (log) & QLIKE & LMAE (log) \\
\midrule
HARQ + $p_t$ & 0.267345 & 0.162259 & 0.402701 \\
Baseline HARQ & 0.279911 & 0.171013 & 0.410499 \\
\bottomrule
\end{tabular}

\vspace{0.4em}
\begin{minipage}{0.90\textwidth}
\footnotesize
\textit{Note:} The table reports out-of-sample volatility-forecasting performance over 2013-08-21 to 2023-05-29. Lower values indicate better forecasting accuracy. $p_t$ denotes the filtered high-volatility probability generated from the residual-based MS--GJR--GARCH model.
\end{minipage}
\end{table}

The regime-augmented specification improves forecasting performance across all three loss functions. The improvement is not confined to a single metric: MSE and LMAE decrease in log space, while QLIKE also improves. This evidence indicates that regime information captures persistent volatility states that are not fully summarized by the standard HARQ structure.

\subsubsection{Forecast calibration}

Table~\ref{tab:mz_results} reports Mincer--Zarnowitz regression results for the baseline HARQ model and the main HARQ + $p_t$ specification.

\begin{table}[H]
\centering
\footnotesize
\setlength{\tabcolsep}{4pt}
\renewcommand{\arraystretch}{1.05}
\caption{Mincer--Zarnowitz regression results}
\label{tab:mz_results}

\begin{tabular}{lcccccc}
\toprule
Model & $N$ & $\alpha$ & $\beta$ & $R^2$ & Joint F-stat. & Joint p-value \\
\midrule
Baseline HARQ & 2326 & -14.9485 & 1.3457 & 0.5215 & 13.6650 & 0.000001 \\
HARQ + $p_t$ & 2326 & -21.2911 & 1.3948 & 0.5398 & 12.3867 & 0.000004 \\
\bottomrule
\end{tabular}

\vspace{0.4em}
\begin{minipage}{0.92\textwidth}
\footnotesize
\textit{Note:} The table reports Mincer--Zarnowitz regressions of realized volatility on model forecasts. The joint test evaluates the null hypothesis $\alpha=0$ and $\beta=1$. HAC/Newey--West standard errors are used with lag length 8.
\end{minipage}
\end{table}

The results show that HARQ + $p_t$ increases the explanatory power of the forecast-realization regression relative to the baseline model. The $R^2$ rises from 0.5215 to 0.5398, indicating that the regime-augmented forecast explains a larger share of realized-volatility variation. The strict calibration null is rejected for both models, implying that neither forecast is perfectly calibrated. Nevertheless, the higher explanatory power of HARQ + $p_t$ suggests that regime information improves forecast quality.

\subsubsection{Conditional predictive accuracy}

To examine whether the forecasting improvement is concentrated in particular volatility states, Table~\ref{tab:conditional_dm} reports conditional Diebold--Mariano tests comparing HARQ + $p_t$ against the baseline HARQ model using QLIKE loss.

\begin{table}[H]
\centering
\footnotesize
\setlength{\tabcolsep}{3.5pt}
\renewcommand{\arraystretch}{1.05}
\caption{Conditional Diebold--Mariano tests for HARQ + $p_t$}
\label{tab:conditional_dm}

\begin{tabular}{llccccc}
\toprule
Split & Regime & Threshold & $N$ & Mean Loss Diff. & DM stat. & p-value \\
\midrule
Median RV & High-vol & 66.7392 & 1163 & 0.015212 & 2.5818 & 0.0049 \\
Median RV & Low-vol & 66.7392 & 1163 & 0.002296 & 1.3890 & 0.0824 \\
\midrule
Top 25\% RV & High-vol & 115.8919 & 582 & 0.022078 & 2.0639 & 0.0195 \\
Top 25\% RV & Low/normal-vol & 115.8919 & 1744 & 0.004308 & 3.0933 & 0.0010 \\
\bottomrule
\end{tabular}

\vspace{0.4em}
\begin{minipage}{0.95\textwidth}
\footnotesize
\textit{Note:} The table reports conditional Diebold--Mariano tests comparing HARQ + $p_t$ against the baseline HARQ model using QLIKE loss. Positive mean loss differences indicate lower QLIKE loss for HARQ + $p_t$. HAC/Newey--West standard errors are used.
\end{minipage}
\end{table}

The conditional tests show that the forecasting gains are strongest during high-volatility periods. Under the median realized-volatility split, the improvement is statistically significant in the high-volatility subsample, while the low-volatility subsample exhibits weaker evidence. Under the top-quartile split, HARQ + $p_t$ significantly outperforms the baseline both in the extreme high-volatility subsample and in the remaining lower-volatility sample. These results suggest that regime augmentation improves volatility forecasting generally, but its largest gains arise during periods of heightened market stress.

Overall, the volatility-forecasting results support the use of HARQ + $p_t$ as the primary first-stage volatility input for the subsequent return-prediction and strategy analysis. Alternative regime specifications and re-estimation-frequency checks are reported in Section~\ref{sec:robustness}.

\subsection{Return prediction results}
\label{sec:return_results}

This subsection evaluates the out-of-sample return-prediction performance of the MS--HARQ--XGBoost framework. The main analysis is conducted over the 3M baseline out-of-sample sample from 2014-12-17 to 2023-05-26, comprising 2,004 observations. This main sample is distinct from the shorter common sample used later for re-estimation-frequency robustness checks.

\subsubsection{Predictive performance}

Table~\ref{tab:return_performance} reports return-prediction performance for the full sample and separately for high- and low-volatility regimes.

\begin{table}[H]
\centering
\footnotesize
\setlength{\tabcolsep}{4pt}
\renewcommand{\arraystretch}{1.05}
\caption{Return prediction performance}
\label{tab:return_performance}

\begin{tabular}{lccccc}
\toprule
Sample & Obs & Corr & Hit Ratio & HAC p-value & MAE \\
\midrule
Full sample & 2004 & 0.0449 & 53.49\% & 0.0009 & 0.0102 \\
High-vol ($p_t > 0.5$) & 384 & -0.0617 & 52.34\% & 0.1928 & 0.0104 \\
Low-vol ($p_t \leq 0.5$) & 1620 & 0.0638 & 53.77\% & 0.0013 & 0.0101 \\
\bottomrule
\end{tabular}

\vspace{0.4em}
\begin{minipage}{0.95\textwidth}
\footnotesize
\textit{Note:} The table reports out-of-sample return-prediction performance for the main 3M baseline sample from 2014-12-17 to 2023-05-26. Corr denotes the Pearson correlation between predicted and realized returns. Hit Ratio denotes directional accuracy. HAC p-values are based on Newey--West adjusted tests of the null hypothesis that directional accuracy equals 50\%. MAE denotes mean absolute error.
\end{minipage}
\end{table}

The results indicate that return predictability is weak but statistically detectable. The full-sample predictive correlation is 0.0449, and the hit ratio is 53.49\%, with a HAC-corrected p-value of 0.0009. The use of HAC correction is important because daily return-prediction errors may exhibit serial dependence and heteroskedasticity.

Predictability is strongly regime dependent. In low-volatility periods, the model achieves a higher correlation of 0.0638 and a statistically significant hit ratio of 53.77\%. In high-volatility regimes, however, the correlation becomes negative and the hit-ratio evidence is statistically insignificant. This pattern suggests that predictive information is concentrated primarily in low-volatility states, while high-volatility states are dominated by unstable shocks and should be treated mainly as exposure-suppression regimes rather than directional-alpha environments.

\subsubsection{Feature importance}

\begin{figure}[H]
\centering
\caption{Average feature importance across walk-forward windows}
\label{fig:feature_importance}
\includegraphics[width=0.85\textwidth]{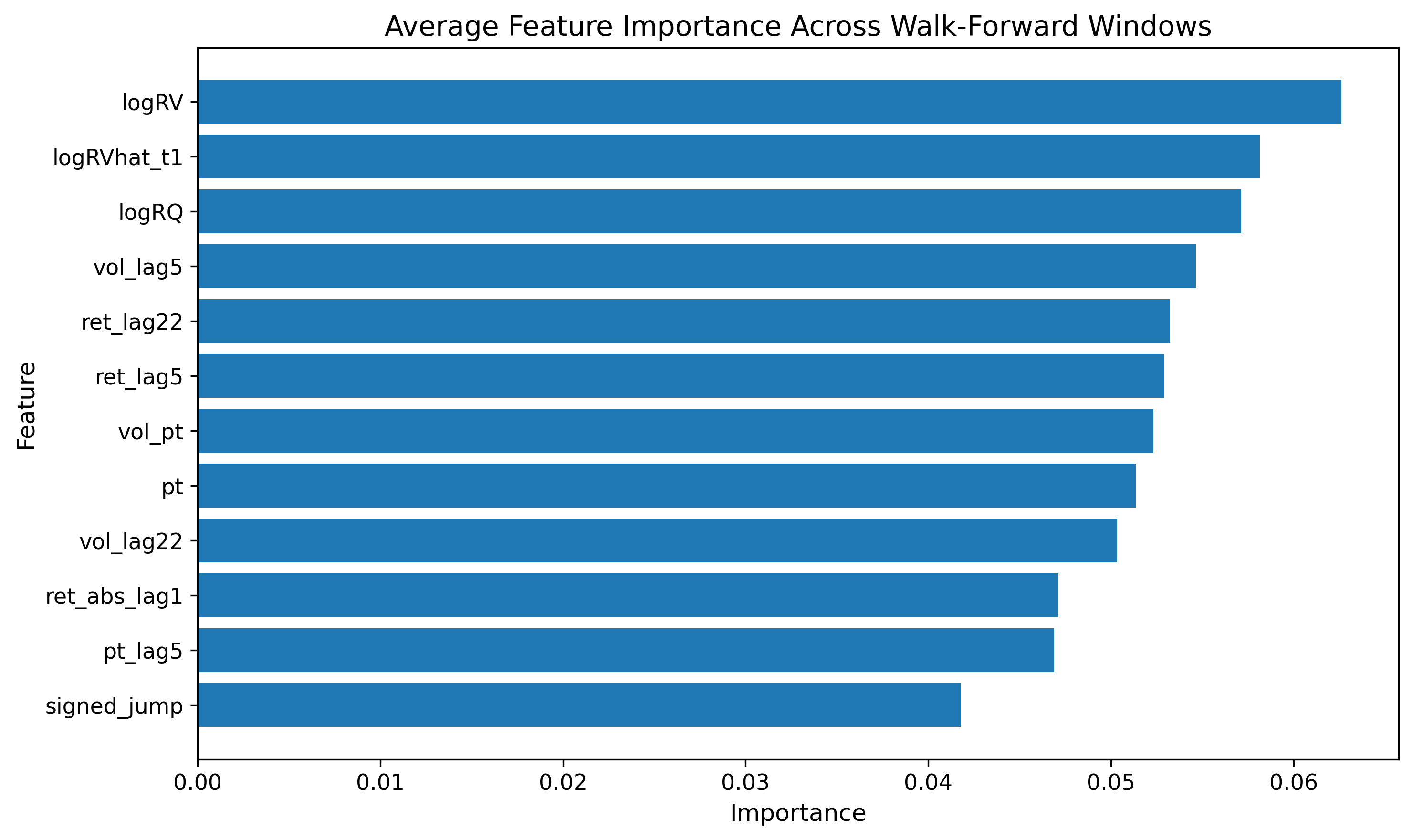}

\figurenote{The figure reports average feature-importance scores across all walk-forward estimation windows based on the XGBoost model. Higher values indicate greater contribution of the corresponding feature to return prediction.}
\end{figure}

Figure~\ref{fig:feature_importance} presents the average feature importance across walk-forward estimation windows.

The feature-importance results reveal a clear hierarchy among predictors. Volatility-related variables, including realized volatility, predicted volatility, and realized quarticity, rank among the most important features. This highlights the dominant role of time-varying risk in driving return predictability.

Return-based predictors play a secondary but non-negligible role. Regime-related variables, including $p_t$ and its interaction terms, provide complementary information but are generally less dominant than volatility features. This pattern is consistent with the later feature-ablation analysis: volatility-related information contributes more directly to raw return prediction, while regime variables are particularly useful for identifying when signals are more or less reliable.

\subsubsection{Predicted versus realized returns}

\begin{figure}[H]
\centering
\caption{Predicted versus realized returns}
\label{fig:scatter_returns}
\includegraphics[width=0.75\textwidth]{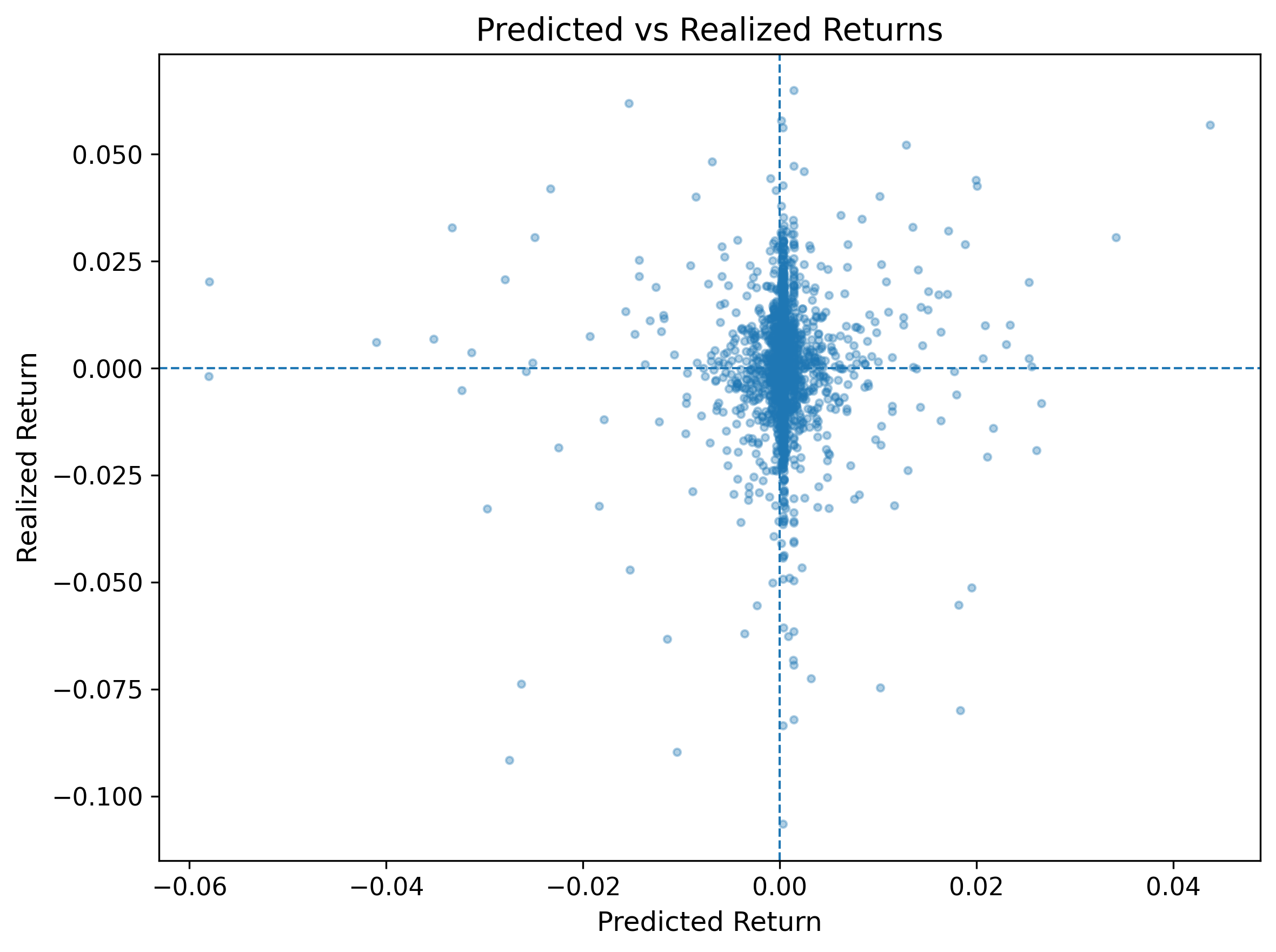}

\figurenote{The figure plots predicted returns against realized returns in the out-of-sample period. The dispersion around the fitted relation reflects the limited predictive accuracy and high noise inherent in return forecasting.}
\end{figure}

Figure~\ref{fig:scatter_returns} plots predicted returns against realized returns over the out-of-sample period.

The scatter plot shows substantial dispersion, with most observations clustered around zero. This confirms the low signal-to-noise ratio inherent in return prediction. Nevertheless, a weak positive association is visible, consistent with the positive full-sample predictive correlation reported in Table~\ref{tab:return_performance}.

\subsubsection{Time-varying predictability}

\begin{figure}[H]
\centering
\caption{Rolling information coefficient}
\label{fig:rolling_ic}
\includegraphics[width=0.85\textwidth]{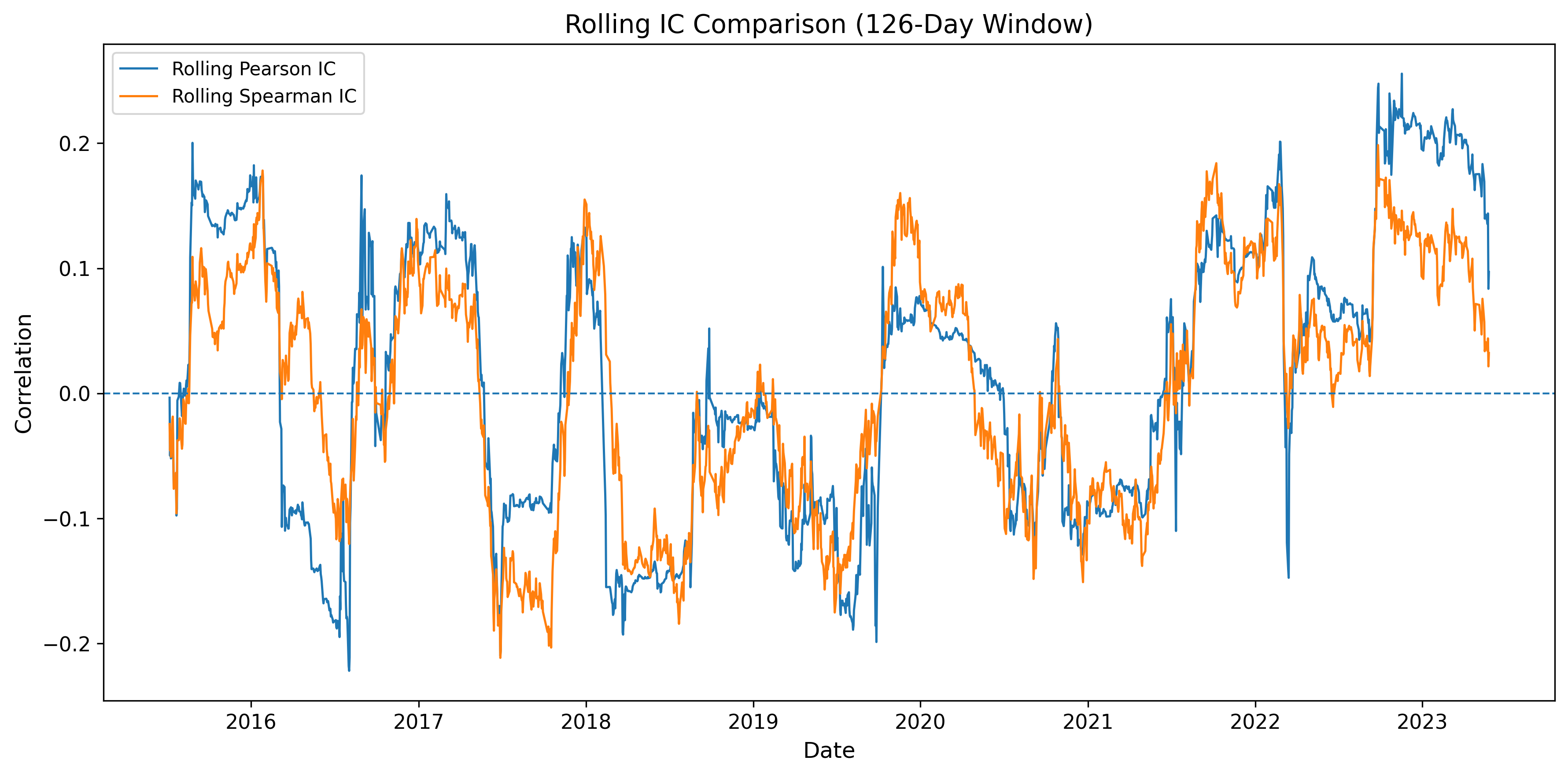}

\figurenote{The figure shows the rolling information coefficient, measuring the correlation between predicted and realized returns over time. The time variation in IC highlights the instability and regime dependence of return predictability.}
\end{figure}

Figure~\ref{fig:rolling_ic} reports the rolling information coefficient of predicted returns.

The rolling IC alternates between positive and negative values, confirming that return predictability is episodic rather than stable. The full-sample predicted return has a weak but positive Pearson IC of 0.0449. However, predictive performance varies considerably across market states.

\begin{figure}[H]
\centering
\caption{Rolling Signal$\times$Risk IC and regime probability}
\label{fig:rolling_ic_pt}
\includegraphics[width=0.90\textwidth]{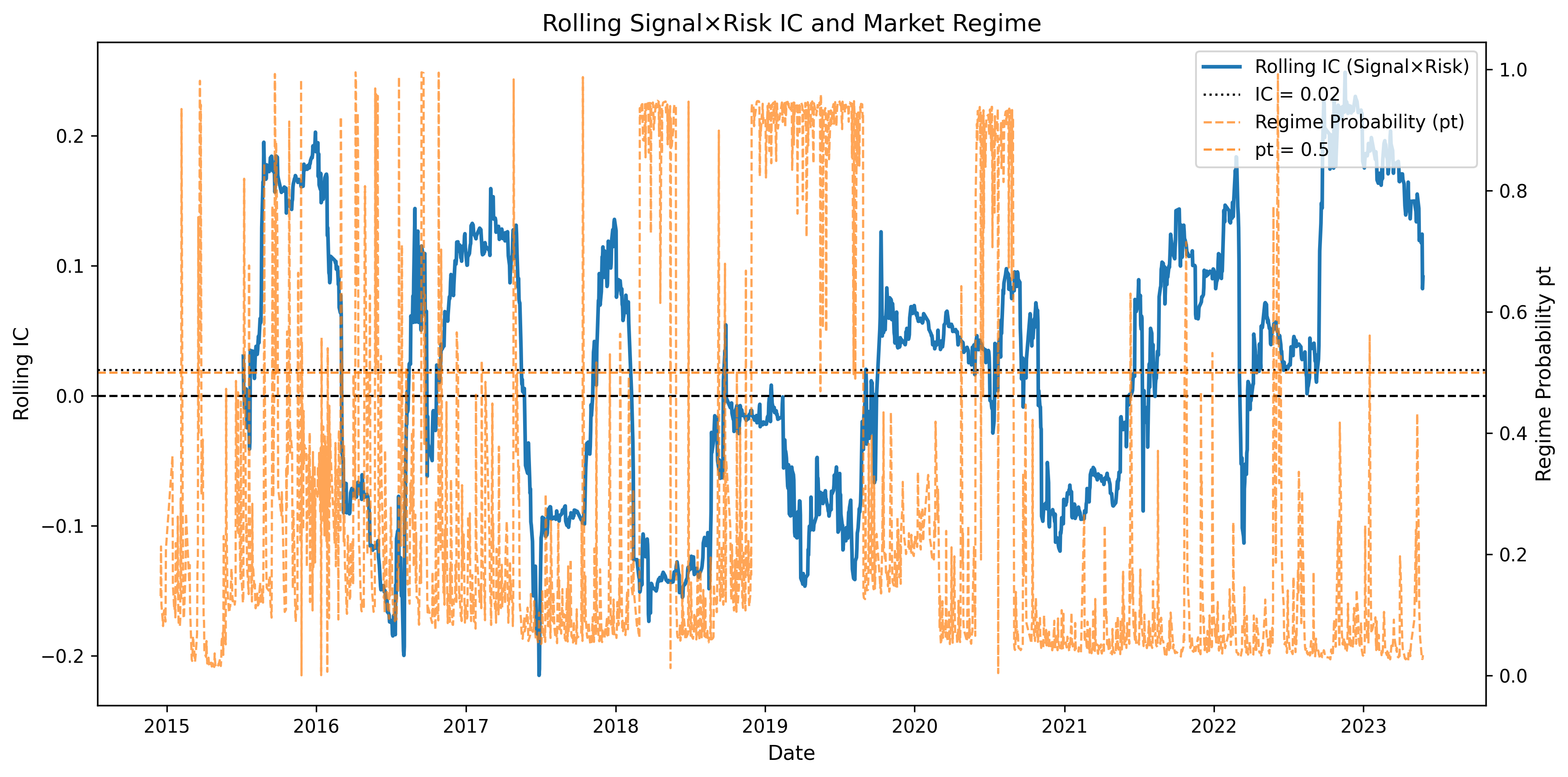}

\figurenote{The figure plots the 126-day rolling information coefficient of the Signal$\times$Risk predictor together with the filtered high-volatility regime probability $p_t$ (orange dashed line, right axis). The horizontal dotted line marks IC = 0.02. Periods of persistently positive IC occur primarily when $p_t$ remains low, whereas elevated $p_t$ is more frequently associated with unstable or negative IC.}
\end{figure}

Figure~\ref{fig:rolling_ic_pt} provides direct evidence that predictive stability is regime dependent.

Positive and persistent rolling IC episodes are concentrated mainly in low-volatility environments. By contrast, high-volatility states are associated with unstable or negative predictive performance. This finding provides empirical support for the Low-vol Gated Weekly Signal$\times$Risk implementation, which concentrates exposure in low-volatility states and reduces participation when predictive reliability deteriorates.

\subsection{Economic value of forecasts}
\label{sec:economic_results}

This subsection evaluates whether weak statistical return predictability can be transformed into economically meaningful portfolio performance. The main implementation is the Low-vol Gated Weekly Signal$\times$Risk strategy described in Section~\ref{sec:strategy}. The strategy is compared with Buy-and-Hold, Momentum Weekly, Vol-managed Buy-and-Hold Weekly, Regime-only Long-Flat, and Regime-only Long-Short benchmarks.

\subsubsection{Out-of-sample equity curves}

\begin{figure}[H]
\centering
\caption{Out-of-sample equity curve}
\label{fig:equity_curve}
\includegraphics[width=0.90\textwidth]{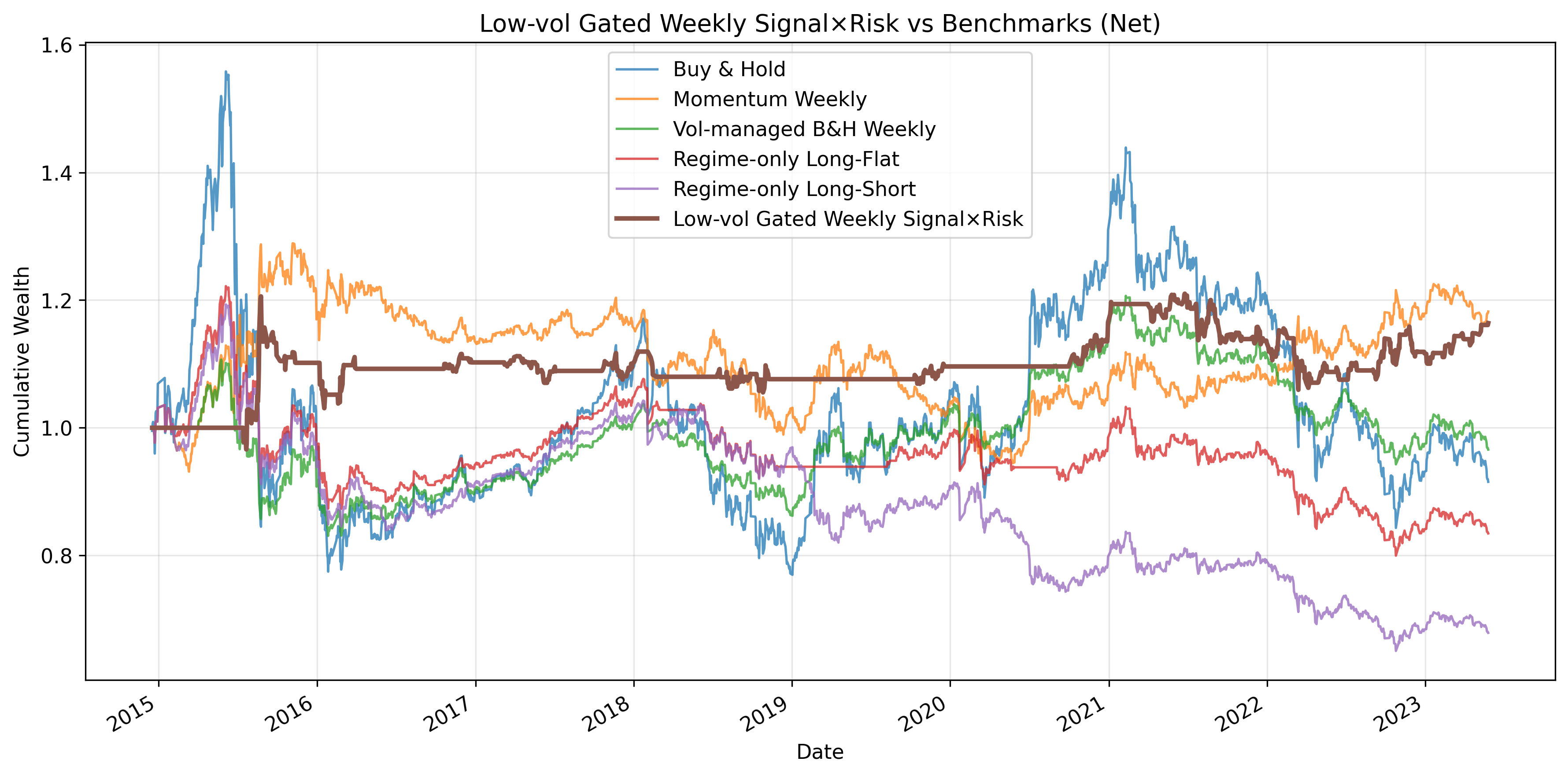}

\figurenote{The figure presents out-of-sample cumulative net wealth from 2014-12-17 to 2023-05-26 for the Low-vol Gated Weekly Signal$\times$Risk strategy and benchmark strategies. All strategy returns are reported net of 5~bp transaction costs.}
\end{figure}

Figure~\ref{fig:equity_curve} presents out-of-sample cumulative net wealth for the main strategy and benchmark strategies.

The Low-vol Gated Weekly Signal$\times$Risk strategy does not maximize cumulative wealth in every subperiod. Instead, its main advantage lies in smoother wealth dynamics and substantially lower drawdowns. Compared with Buy-and-Hold, which experiences strong boom-bust cycles, the proposed strategy maintains a more stable equity trajectory, particularly during turbulent periods such as the 2015--2016 crash.

Relative to Momentum Weekly, the proposed strategy generates slightly lower terminal wealth but does so with materially lower volatility and shallower drawdowns. This confirms that the strategy is primarily defensive and risk-controlled rather than return-maximizing.

\subsubsection{Strategy performance}

Table~\ref{tab:strategy_performance_supervisor} reports net performance metrics for the main strategy and benchmarks.

\begin{table}[H]
\centering
\caption{Net performance of main strategy and benchmarks}
\label{tab:strategy_performance_supervisor}
\begin{threeparttable}
\scriptsize
\setlength{\tabcolsep}{2.4pt}
\resizebox{\textwidth}{!}{
\begin{tabular}{lcccccccccccc}
\toprule
Strategy & ARC & ASD & MD & Sharpe & Sortino & IR$^{**}$ & MLD & IR$^{***}$ & No. trades & \%Long & \%Short & \%Neutral \\
\midrule
Buy \& Hold & -1.11 & 23.38 & -50.60 & -0.047 & -0.062 & -- & 1899 & -- & 1 & 100.00 & 0.00 & 0.00 \\
Momentum Weekly & 2.13 & 11.68 & -26.71 & 0.182 & 0.258 & 0.042 & 1798 & 0.065 & 92 & 54.39 & 45.51 & 0.10 \\
Vol-managed B\&H Weekly & -0.44 & 10.04 & -24.94 & -0.043 & -0.057 & -0.110 & 1315 & -0.166 & 272 & 96.96 & 0.00 & 3.04 \\
Regime-only Long-Flat & -2.25 & 10.69 & -34.55 & -0.210 & -0.272 & -0.235 & 1899 & -0.342 & 49 & 79.84 & 0.00 & 20.16 \\
Regime-only Long-Short & -4.75 & 11.69 & -45.49 & -0.407 & -0.528 & -0.328 & 1899 & -0.463 & 49 & 79.84 & 20.06 & 0.10 \\
Low-vol Gated Weekly Signal$\times$Risk & 1.93 & 7.56 & -14.52 & 0.255 & 0.394 & 0.021 & 1384 & 0.034 & 163 & 20.66 & 12.57 & 66.77 \\
\bottomrule
\end{tabular}
}
\begin{tablenotes}[para,flushleft]
\scriptsize
\item Note: The table reports net performance after 5~bp transaction costs over the main 3M out-of-sample sample from 2014-12-17 to 2023-05-26. ARC, ASD, MD, \%Long, \%Short, and \%Neutral are reported in percentage terms. ARC denotes annualized return compounded, ASD denotes annualized standard deviation, and MD denotes maximum drawdown. IR$^{**}$ denotes the information ratio relative to Buy-and-Hold based on tracking error, while IR$^{***}$ uses downside tracking error. MLD denotes maximum loss duration in trading days.
\end{tablenotes}
\end{threeparttable}
\end{table}

The main strategy delivers the lowest maximum drawdown among the evaluated strategies. Net of 5~bp transaction costs, it achieves an ARC of 1.93\%, an ASD of 7.56\%, a Sharpe ratio of 0.255, and a Sortino ratio of 0.394. Although Momentum Weekly generates a slightly higher annualized return, it does so with substantially higher volatility and drawdown. The proposed strategy has a maximum drawdown of -14.52\%, compared with -50.60\% for Buy-and-Hold and -26.71\% for Momentum Weekly.

The exposure diagnostics show that the strategy is highly selective. It is long on 20.66\% of days, short on 12.57\% of days, and neutral on 66.77\% of days. This confirms that the strategy is not designed to remain continuously invested, but instead participates only when the predictive signal is sufficiently strong and market conditions are favorable.

The benchmark-relative information ratios are positive but small, indicating that active performance relative to Buy-and-Hold is economically positive but modest when measured against tracking-error risk. The main contribution of the strategy is therefore better interpreted as risk reduction and downside protection rather than large benchmark-relative alpha.

\subsubsection{Bootstrap inference}

Table~\ref{tab:sharpe_bootstrap} reports circular block bootstrap tests for strategy Sharpe ratios.

\begin{table}[htbp]
\centering
\footnotesize
\setlength{\tabcolsep}{4pt}
\renewcommand{\arraystretch}{1.05}
\caption{Circular block bootstrap tests for Sharpe ratios}
\label{tab:sharpe_bootstrap}

\begin{tabular}{lcccc}
\toprule
Strategy & Observed Sharpe & CI Lower & CI Upper & p-value \\
\midrule
Buy \& Hold & -0.047 & -0.700 & 0.649 & 0.413 \\
Momentum Weekly & 0.182 & -0.666 & 0.579 & 0.225 \\
Vol-managed B\&H Weekly & -0.043 & -0.716 & 0.720 & 0.482 \\
Regime-only Long-Flat & -0.210 & -0.676 & 0.684 & 0.679 \\
Regime-only Long-Short & -0.407 & -0.708 & 0.675 & 0.846 \\
Low-vol Gated Weekly Signal$\times$Risk & 0.255 & -0.683 & 0.599 & 0.182 \\
\bottomrule
\end{tabular}

\vspace{0.4em}
\begin{minipage}{0.95\textwidth}
\footnotesize
\textit{Note:} The table reports circular block bootstrap tests of the null hypothesis that the Sharpe ratio is less than or equal to zero. The bootstrap procedure uses 5,000 replications and a block length of 20 trading days.
\end{minipage}
\end{table}

The bootstrap results provide a more cautious interpretation of economic performance. The proposed strategy has the highest observed Sharpe ratio among the evaluated strategies, but the p-value for testing whether the Sharpe ratio is greater than zero is 0.182. The confidence interval is wide, reflecting substantial uncertainty in Sharpe-ratio estimation.

Therefore, the evidence is economically suggestive but not statistically strong. The strategy should be interpreted as defensive and economically promising, rather than statistically dominant.

\subsubsection{Sub-period and event-window performance}

To examine when the strategy creates value, performance is evaluated across annual subperiods and major market-event windows.

\begin{figure}[H]
\centering
\caption{Annual return comparison}
\label{fig:annual_return_comparison}
\includegraphics[width=0.90\textwidth]{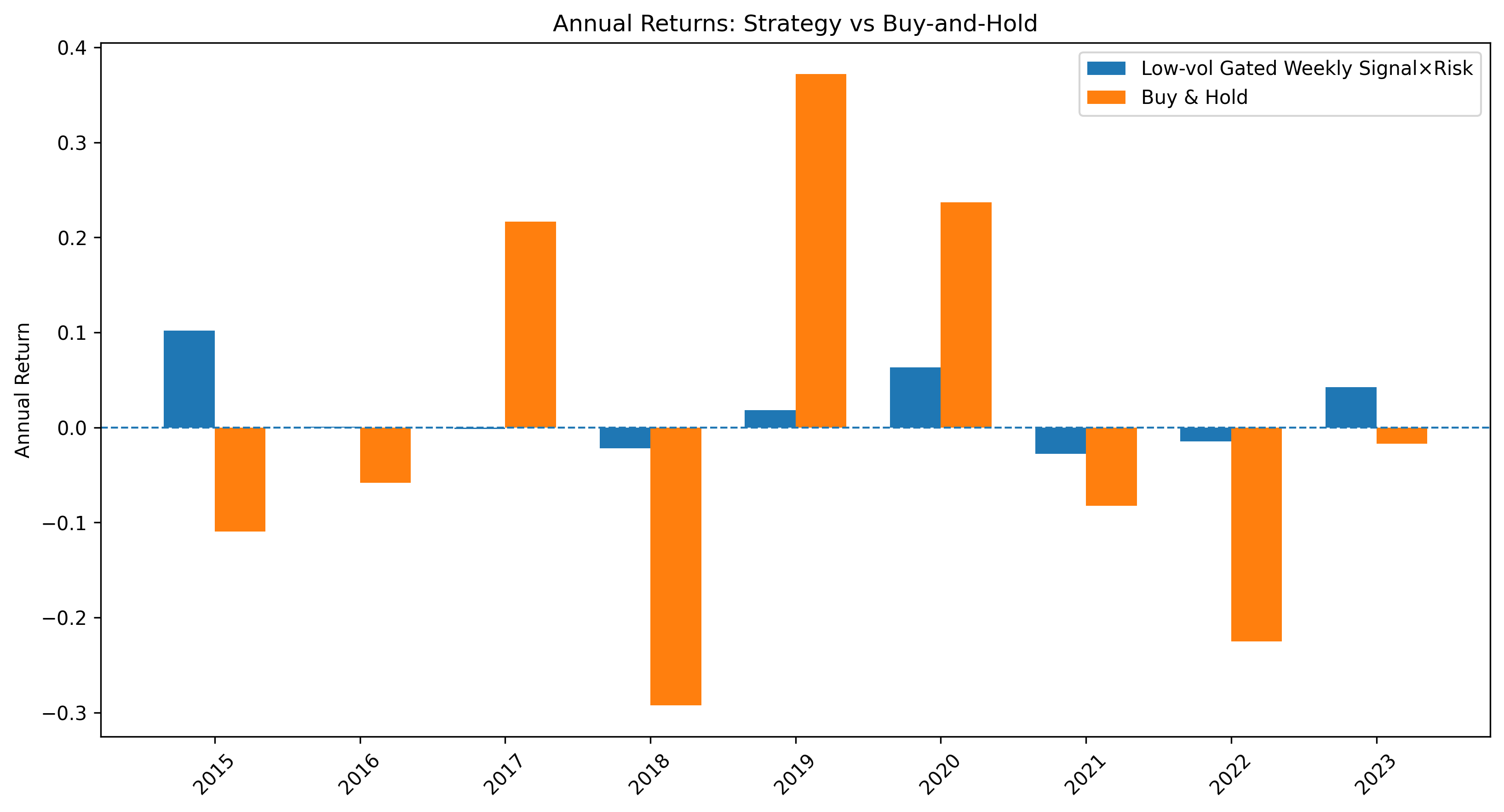}

\figurenote{The figure compares annual returns of the Low-vol Gated Weekly Signal$\times$Risk strategy and Buy-and-Hold. The strategy tends to outperform during adverse or turbulent market periods, while underperforming in strong bull-market years.}
\end{figure}

Figure~\ref{fig:annual_return_comparison} compares annual returns of the proposed strategy and Buy-and-Hold.

The strategy outperforms Buy-and-Hold in 6 out of 9 calendar-year periods, with an average annual return advantage of approximately 1.33 percentage points. However, outperformance is uneven and concentrated in adverse or turbulent periods. The strategy performs particularly well in 2015, 2018, 2022, and the partial 2023 period, while underperforming in strong bull-market years such as 2017 and 2019.

\begin{figure}[H]
\centering
\caption{Annual maximum drawdown comparison}
\label{fig:annual_maxdd_comparison}
\includegraphics[width=0.90\textwidth]{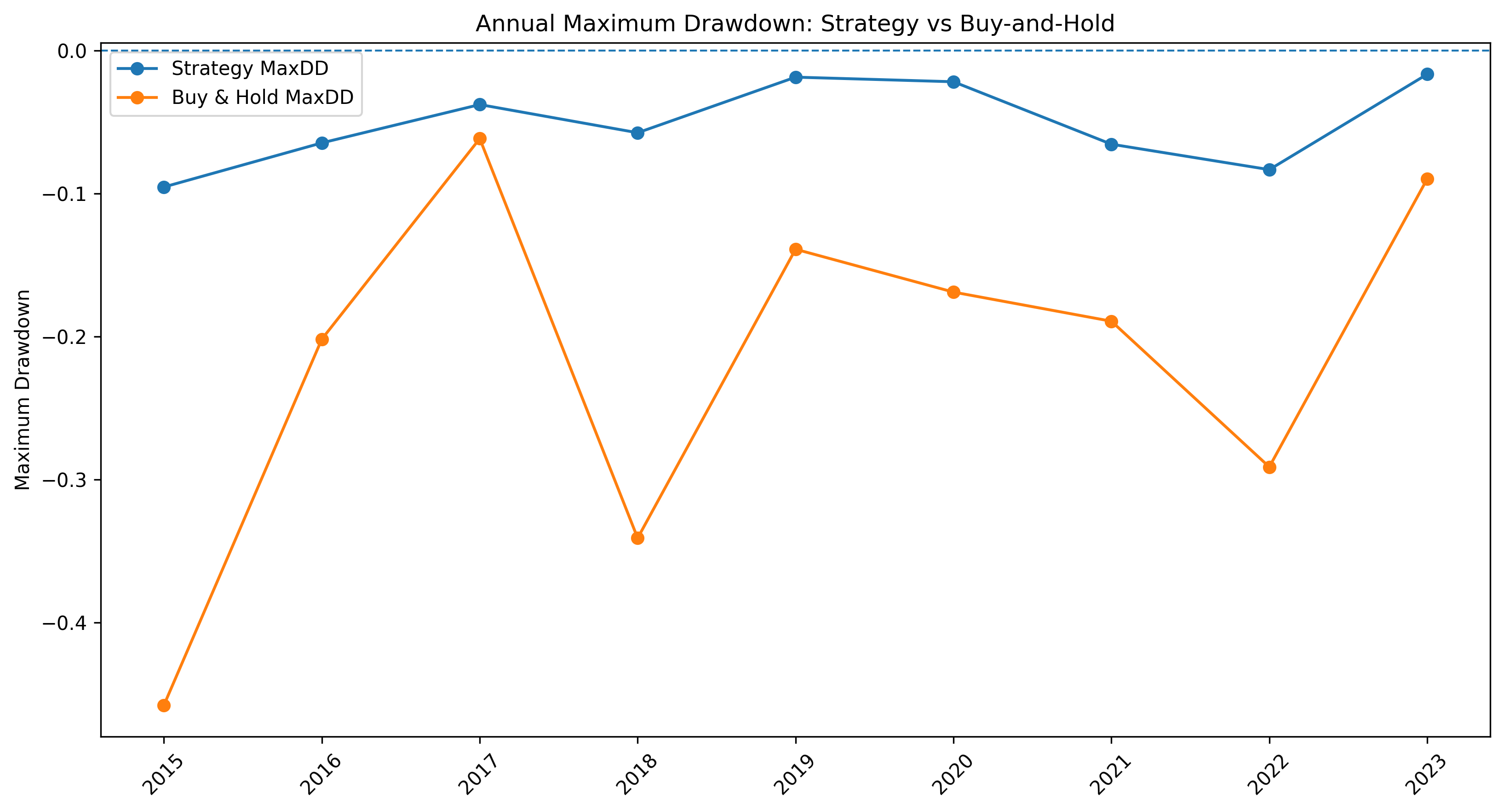}

\figurenote{The figure compares annual maximum drawdowns of the Low-vol Gated Weekly Signal$\times$Risk strategy and Buy-and-Hold. The strategy consistently exhibits smaller drawdowns, indicating stronger downside protection.}
\end{figure}

Figure~\ref{fig:annual_maxdd_comparison} compares annual maximum drawdowns.

The strategy consistently reduces drawdown severity relative to Buy-and-Hold. Even in years when it underperforms in terms of raw return, drawdown reduction remains substantial. This confirms that the strategy's primary value lies in capital preservation and downside-risk control.

Table~\ref{tab:event_window_performance} reports performance across major event windows.

\begin{table}[H]
\centering
\footnotesize
\setlength{\tabcolsep}{3pt}
\renewcommand{\arraystretch}{1.05}
\caption{Event-window performance}
\label{tab:event_window_performance}

\resizebox{\textwidth}{!}{%
\begin{tabular}{lrrrrrr}
\toprule
Event window & Strategy Ret. & B\&H Ret. & Diff. & Strategy Sharpe & B\&H Sharpe & Strategy / B\&H MaxDD \\
\midrule
2015--16 Crash & 10.24\% & -39.38\% & 49.62\% & 0.546 & -0.855 & -14.52\% / -50.29\% \\
2018--19 Trade War & -0.40\% & -2.91\% & 2.51\% & -0.043 & -0.071 & -5.74\% / -34.26\% \\
2020 COVID & 6.32\% & 23.71\% & -17.38\% & 1.581 & 1.077 & -2.17\% / -16.88\% \\
Post-COVID 2021--23 & -0.12\% & -30.11\% & 29.99\% & -0.006 & -0.793 & -12.44\% / -41.41\% \\
\bottomrule
\end{tabular}%
}

\vspace{0.4em}
\begin{minipage}{0.95\textwidth}
\footnotesize
\textit{Note:} The table reports cumulative event-window returns, Sharpe ratios, and maximum drawdowns for the Low-vol Gated Weekly Signal$\times$Risk strategy and Buy-and-Hold. The evidence shows that the strategy's economic value is concentrated primarily in downside protection during adverse market environments rather than upside capture during strong bull markets.
\end{minipage}
\end{table}

The event-window evidence confirms the defensive nature of the strategy. During the 2015--2016 Chinese equity-market crash, the strategy generates a cumulative return of 10.24\%, compared with -39.38\% for Buy-and-Hold. During the post-COVID 2021--2023 period, the strategy approximately preserves capital, while Buy-and-Hold loses 30.11\%. By contrast, during the 2020 COVID rebound, Buy-and-Hold generates stronger raw returns, although the strategy still delivers substantially smaller drawdowns.

Overall, the empirical evidence indicates that the Low-vol Gated Weekly Signal$\times$Risk strategy creates value primarily through defensive exposure management, drawdown control, and selective participation. Its economic value is concentrated in adverse or turbulent market environments rather than in unconditional return maximization.

\subsection{Summary of empirical results}
\label{sec:results_summary}

The empirical results reveal three main findings.

First, regime information improves volatility forecasting. HARQ + $p_t$ outperforms the baseline HARQ model across MSE, QLIKE, and LMAE, exhibits higher explanatory power in Mincer--Zarnowitz regressions, and delivers especially strong improvements during elevated-volatility periods.

Second, return predictability is weak but state dependent. Full-sample predictive correlation is positive but small, while directional accuracy is statistically significant after HAC correction. Predictive performance is materially stronger in low-volatility states and deteriorates in high-volatility regimes.

Third, weak return forecasts can generate economic value only when combined with disciplined implementation. The Low-vol Gated Weekly Signal$\times$Risk strategy improves Sharpe and Sortino ratios relative to Buy-and-Hold, reduces annualized volatility and maximum drawdown, and remains neutral during a large fraction of the sample. However, bootstrap inference indicates that the evidence is economically meaningful but statistically modest.

Overall, the findings support the view that volatility and regime information are most valuable as risk-management and signal-reliability tools rather than as sources of strong unconditional return predictability.

\section{Robustness and diagnostics}
\label{sec:robustness}

This section evaluates the robustness of the main findings along several dimensions. While the previous section focused on the baseline MS--HARQ--XGBoost framework and the Low-vol Gated Weekly Signal$\times$Risk strategy, the analyses reported here examine whether the conclusions remain valid under alternative volatility specifications, re-estimation frequencies, predictor sets, implementation rules, and parameter choices.

The objective is not to identify an ex post best-performing configuration, but to assess whether the central conclusions are sensitive to reasonable modeling and implementation alternatives.

\subsection{Volatility-model robustness}
\label{subsec:vol_model_robustness}

This subsection evaluates whether the volatility-forecasting gains documented in the main results section are robust to alternative representations of regime dependence.

The baseline specification augments the HARQ model with the filtered high-volatility probability $p_t$. Two alternative specifications are considered. The first replaces the filtered probability with a hard regime indicator $D_t$, while the second introduces a parsimonious regime-dependent interaction term involving daily realized volatility.

All specifications are evaluated using the same walk-forward framework and the identical out-of-sample window from 2013-08-21 to 2023-05-29 ($N=2326$). The objective is to determine whether the documented forecasting gains arise from the specific choice of regime variable or from the broader incorporation of regime information itself.

\begin{table}[!htbp]
\centering
\footnotesize
\setlength{\tabcolsep}{4pt}
\renewcommand{\arraystretch}{1.05}
\caption{Mincer--Zarnowitz regressions across volatility model specifications}
\label{tab:vol_spec_mz_robustness}

\begin{tabular}{lcccccc}
\toprule
Model & Obs & $\alpha$ & $\beta$ & $R^2$ & Joint F-stat. & Joint p-value \\
\midrule
Baseline HARQ & 2326 & -14.9485 & 1.3457 & 0.5215 & 13.6650 & 0.000001 \\
HARQ + $p_t$ & 2326 & -21.2911 & 1.3948 & 0.5398 & 12.3867 & 0.000004 \\
HARQ + $D_t$ & 2326 & -21.1918 & 1.3959 & 0.5396 & 12.6563 & 0.000003 \\
Single interaction & 2326 & -20.1016 & 1.3857 & 0.5336 & 12.6410 & 0.000003 \\
\bottomrule
\end{tabular}

\tablenote{The table reports Mincer--Zarnowitz regressions of realized volatility on model forecasts. The joint test evaluates $H_0:\alpha=0,\beta=1$. HAC/Newey--West standard errors are used with lag length 8. Higher $R^2$ indicates stronger forecast-realization alignment.}
\end{table}

Table~\ref{tab:vol_spec_mz_robustness} shows that all regime-augmented specifications improve forecast-realization alignment relative to the baseline HARQ model. The increase in $R^2$ is consistent across alternative regime representations. The HARQ + $p_t$ and HARQ + $D_t$ specifications produce nearly identical calibration performance, while the single-interaction model remains slightly weaker but still improves upon the baseline.

Although the joint null of perfect calibration is rejected for all specifications, the results indicate that incorporating regime information improves the empirical relationship between forecasted and realized volatility.

\begin{table}[!htbp]
\centering
\footnotesize
\setlength{\tabcolsep}{2.5pt}
\renewcommand{\arraystretch}{1.05}
\caption{Conditional Diebold--Mariano tests across volatility model specifications}
\label{tab:vol_spec_conditional_dm}

\begin{tabular}{llccccc}
\toprule
Split & Model & Regime & Obs & Mean Loss Diff. & DM stat. & p-value \\
\midrule
Median RV & HARQ + $p_t$ & High-vol & 1163 & 0.015212 & 2.5818 & 0.0049 \\
Median RV & HARQ + $p_t$ & Low-vol & 1163 & 0.002296 & 1.3890 & 0.0824 \\
Median RV & HARQ + $D_t$ & High-vol & 1163 & 0.015156 & 2.6027 & 0.0046 \\
Median RV & HARQ + $D_t$ & Low-vol & 1163 & 0.002645 & 1.6546 & 0.0490 \\
Median RV & Single interaction & High-vol & 1163 & 0.014986 & 2.5579 & 0.0053 \\
Median RV & Single interaction & Low-vol & 1163 & 0.002185 & 1.3593 & 0.0870 \\
\midrule
Top 25\% RV & HARQ + $p_t$ & High-vol & 582 & 0.022078 & 2.0639 & 0.0195 \\
Top 25\% RV & HARQ + $p_t$ & Low/normal-vol & 1744 & 0.004308 & 3.0933 & 0.0010 \\
Top 25\% RV & HARQ + $D_t$ & High-vol & 582 & 0.021832 & 2.0654 & 0.0194 \\
Top 25\% RV & HARQ + $D_t$ & Low/normal-vol & 1744 & 0.004585 & 3.3405 & 0.0004 \\
Top 25\% RV & Single interaction & High-vol & 582 & 0.021068 & 1.9619 & 0.0249 \\
Top 25\% RV & Single interaction & Low/normal-vol & 1744 & 0.004420 & 3.2154 & 0.0007 \\
\bottomrule
\end{tabular}

\tablenote{The table reports conditional Diebold--Mariano tests comparing each regime-augmented specification against the baseline HARQ model using QLIKE loss. Positive mean loss differences indicate lower QLIKE loss for the regime-augmented model. HAC/Newey--West standard errors are used.}
\end{table}

Table~\ref{tab:vol_spec_conditional_dm} reports conditional Diebold--Mariano tests based on QLIKE loss. The results indicate that regime augmentation improves volatility forecasts across all alternative specifications. Forecasting gains are strongest during elevated-volatility periods, where all regime-augmented models significantly outperform the baseline HARQ specification. Improvements remain positive in lower-volatility environments, although the magnitude of the gains is smaller.

Taken together, the evidence suggests that the forecasting improvements documented in the main analysis are robust to alternative representations of regime dependence. The primary benefit arises from incorporating regime information itself rather than from a particular transformation of the regime variable.

\subsection{Re-estimation-frequency robustness in volatility forecasting}
\label{subsec:frequency_robustness}

This subsection evaluates whether volatility-forecasting performance is sensitive to the frequency at which model parameters are updated.

To ensure direct comparability, all specifications are evaluated on the common out-of-sample period from 2013-11-19 to 2023-05-29 ($N=2269$). Monthly (1M), quarterly (3M), and semi-annual (6M) re-estimation frequencies are considered.

\begin{table}[!htbp]
\centering
\scriptsize
\setlength{\tabcolsep}{2.5pt}
\renewcommand{\arraystretch}{1.05}
\caption{Volatility forecasting robustness across re-estimation frequencies and model specifications}
\label{tab:vol_frequency_spec_robustness}

\resizebox{\textwidth}{!}{%
\begin{tabular}{llcccccc}
\toprule
Frequency & Model & Obs & MSE (log) & QLIKE & LMAE (log) & DM stat. & DM p-value \\
\midrule
1M & Baseline HARQ & 2269 & 0.282423 & 0.173066 & 0.412373 & -- & -- \\
1M & HARQ + $D_t$ & 2269 & 0.268566 & 0.166198 & 0.402499 & 2.4481 & 0.0072 \\
1M & HARQ + $p_t$ & 2269 & 0.267902 & 0.164842 & 0.402272 & 3.0203 & 0.0013 \\
1M & Single interaction & 2269 & 0.269597 & 0.166399 & 0.403311 & 2.5457 & 0.0055 \\
\midrule
3M & Baseline HARQ & 2269 & 0.282564 & 0.173071 & 0.412549 & -- & -- \\
3M & HARQ + $D_t$ & 2269 & 0.268991 & 0.163992 & 0.403604 & 3.2436 & 0.0006 \\
3M & HARQ + $p_t$ & 2269 & 0.269139 & 0.164103 & 0.403741 & 3.1760 & 0.0007 \\
3M & Single interaction & 2269 & 0.269526 & 0.164318 & 0.403935 & 3.1052 & 0.0010 \\
\midrule
6M & Baseline HARQ & 2269 & 0.282690 & 0.173060 & 0.412713 & -- & -- \\
6M & HARQ + $D_t$ & 2269 & 0.269096 & 0.163354 & 0.404060 & 3.3262 & 0.0004 \\
6M & HARQ + $p_t$ & 2269 & 0.270101 & 0.164024 & 0.405153 & 3.1477 & 0.0008 \\
6M & Single interaction & 2269 & 0.269778 & 0.163857 & 0.404845 & 3.1907 & 0.0007 \\
\bottomrule
\end{tabular}%
}

\tablenote{The table reports volatility-forecasting performance across re-estimation frequencies and model specifications on the common out-of-sample period from 2013-11-19 to 2023-05-29 ($N=2269$). DM statistics compare each regime-augmented specification against the baseline HARQ model using QLIKE loss and HAC/Newey--West standard errors. Lower MSE, QLIKE, and LMAE indicate better forecasting performance.}
\end{table}

Table~\ref{tab:vol_frequency_spec_robustness} demonstrates that the forecasting gains associated with regime augmentation remain robust across re-estimation frequencies. For every frequency considered, all regime-augmented specifications outperform the baseline HARQ model across MSE, QLIKE, and LMAE.

The QLIKE-based Diebold--Mariano tests are statistically significant for all regime-augmented models, with p-values below 1\% throughout. Although no single frequency dominates across every evaluation metric, the differences between 1M, 3M, and 6M are relatively small compared with the consistent improvements generated by regime information.

Accordingly, the quarterly specification is retained as the baseline because it provides a reasonable balance between adaptability and parameter stability while preserving the main forecasting conclusions.

\subsection{Re-estimation-frequency robustness in return prediction and strategy performance}
\label{subsec:robust_return_frequency}

This subsection examines whether return predictability and strategy performance are robust to the re-estimation frequency. Unlike volatility forecasting, return prediction has a much lower signal-to-noise ratio and is therefore more sensitive to how frequently model parameters are updated. To ensure comparability, the 1M, 3M, and 6M specifications are evaluated on the same common out-of-sample window from 2015-03-30 to 2023-05-26, comprising 1,947 observations.

\begin{table}[htbp]
\centering
\footnotesize
\setlength{\tabcolsep}{3pt}
\renewcommand{\arraystretch}{1.05}
\caption{Return prediction performance across re-estimation frequencies}
\label{tab:return_frequency_robustness}

\resizebox{0.90\textwidth}{!}{%
\begin{tabular}{llccccc}
\toprule
Frequency & Regime & Obs & Corr & MAE & Hit Ratio & HAC p-value \\
\midrule
1M & Full sample & 1947 & 0.0143 & 0.009881 & 51.46\% & 0.1727 \\
1M & High-vol ($p_t>0.5$) & 423 & -0.1016 & 0.011156 & 50.12\% & 0.9574 \\
1M & Low-vol ($p_t\leq 0.5$) & 1524 & 0.0395 & 0.009527 & 51.84\% & 0.1351 \\
\midrule
3M & Full sample & 1947 & 0.0381 & 0.009998 & 53.52\% & 0.0010 \\
3M & High-vol ($p_t>0.5$) & 378 & -0.0867 & 0.010306 & 51.85\% & 0.4167 \\
3M & Low-vol ($p_t\leq 0.5$) & 1569 & 0.0584 & 0.009924 & 53.92\% & 0.0014 \\
\midrule
6M & Full sample & 1947 & -0.0202 & 0.010061 & 51.57\% & 0.1389 \\
6M & High-vol ($p_t>0.5$) & 365 & -0.0532 & 0.011845 & 50.68\% & 0.7678 \\
6M & Low-vol ($p_t\leq 0.5$) & 1582 & 0.0052 & 0.009649 & 51.77\% & 0.1525 \\
\bottomrule
\end{tabular}%
}

\tablenote{The table reports return-prediction performance across re-estimation frequencies on the common out-of-sample sample. HAC p-values are based on Newey--West standard errors and constitute the primary statistical inference. Binomial p-values are omitted for conciseness.}
\end{table}

The return-prediction results support the 3M specification as the most statistically defensible frequency in this comparison. It achieves the highest full-sample correlation, the highest hit ratio, and statistically significant HAC-corrected directional accuracy. The same conclusion holds within the low-volatility regime.

By contrast, neither the 1M nor the 6M specification achieves statistical significance under HAC inference. Across all frequencies, high-volatility regimes remain difficult to predict. Correlations are negative in high-volatility states for all three frequencies, while HAC-corrected hit-ratio tests remain insignificant. These findings reinforce the interpretation that return predictability is concentrated in low-volatility states and provide additional support for the low-volatility gating mechanism adopted in the main strategy.

\begin{table}[htbp]
\centering
\scriptsize
\setlength{\tabcolsep}{3pt}
\renewcommand{\arraystretch}{1.05}
\caption{Economic performance across re-estimation frequencies and benchmarks}
\label{tab:frequency_strategy_robustness}

\resizebox{\textwidth}{!}{%
\begin{tabular}{llccccc}
\toprule
Frequency & Strategy & Ann. Return & Ann. Vol & Sharpe & MaxDD & Break-even Cost (bp) \\
\midrule
1M & Buy \& Hold & -3.47\% & 23.10\% & -0.150 & -50.60\% & 0.00 \\
1M & Vol-managed B\&H & -0.10\% & 8.55\% & -0.012 & -24.46\% & 3.55 \\
1M & Momentum & 1.84\% & 11.56\% & 0.160 & -25.60\% & 13.22 \\
1M & Regime-only LS & 3.32\% & 11.55\% & 0.287 & -22.32\% & 26.64 \\
1M & Regime-only Long-only & 1.40\% & 11.95\% & 0.117 & -25.72\% & 20.29 \\
1M & Main Signal$\times$Risk & -3.00\% & 7.92\% & -0.378 & -30.34\% & 0.00 \\
\midrule
3M & Buy \& Hold & -3.47\% & 23.10\% & -0.150 & -50.60\% & 0.00 \\
3M & Vol-managed B\&H & -0.09\% & 8.56\% & -0.010 & -24.50\% & 3.73 \\
3M & Momentum & 1.84\% & 11.56\% & 0.160 & -25.60\% & 13.22 \\
3M & Regime-only LS & -9.34\% & 11.54\% & -0.809 & -60.61\% & 0.00 \\
3M & Regime-only Long-only & -6.32\% & 12.55\% & -0.504 & -50.95\% & 0.00 \\
3M & Main Signal$\times$Risk & 1.93\% & 8.51\% & 0.226 & -22.64\% & 8.96 \\
\midrule
6M & Buy \& Hold & -3.47\% & 23.10\% & -0.150 & -50.60\% & 0.00 \\
6M & Vol-managed B\&H & -0.14\% & 8.55\% & -0.017 & -24.58\% & 2.86 \\
6M & Momentum & 1.84\% & 11.56\% & 0.160 & -25.60\% & 13.22 \\
6M & Regime-only LS & -4.51\% & 11.56\% & -0.390 & -39.94\% & 0.00 \\
6M & Regime-only Long-only & -3.33\% & 12.32\% & -0.270 & -34.60\% & 0.00 \\
6M & Main Signal$\times$Risk & -4.43\% & 6.99\% & -0.634 & -36.55\% & 0.00 \\
\bottomrule
\end{tabular}%
}

\tablenote{The table reports net performance under a 5~bp transaction cost over the common out-of-sample period from 2015-03-30 to 2023-05-26. Break-even cost denotes the transaction-cost level at which the strategy Sharpe ratio falls to zero. The main strategy performance in this table is not directly comparable with the main 3M baseline reported in the Results section because the present analysis uses the shorter common out-of-sample window required for cross-frequency comparison.}
\end{table}

The economic results are broadly consistent with the statistical evidence. Under 5~bp transaction costs, the 3M Main Signal$\times$Risk strategy is the only main-strategy frequency that delivers both positive annualized return and a positive Sharpe ratio. Specifically, the strategy achieves an annualized return of 1.93\%, a Sharpe ratio of 0.226, and a maximum drawdown of -22.64\%.

By contrast, the 1M main strategy records a Sharpe ratio of -0.378, while the 6M main strategy produces a Sharpe ratio of -0.634. These results suggest that excessively frequent updating may increase estimation noise and turnover, whereas infrequent updating may reduce the model's ability to adapt to changing market conditions.

The benchmark comparison shows that some simpler strategies can perform well in particular configurations. For example, the 1M regime-only long-short strategy and the momentum benchmark generate positive Sharpe ratios. However, these benchmark results are not stable across frequencies and reflect different economic mechanisms. The 3M Main Signal$\times$Risk strategy remains the only main implementation that is economically viable under 5~bp transaction costs on the common out-of-sample sample.

\begin{table}[H]
\centering
\footnotesize
\setlength{\tabcolsep}{5pt}
\renewcommand{\arraystretch}{1.05}
\caption{Transaction-cost sensitivity of the main strategy across re-estimation frequencies}
\label{tab:frequency_cost_sensitivity}

\begin{tabular}{lcccc}
\toprule
Frequency & Cost & Ann. Return & Sharpe & MaxDD \\
\midrule
1M & 0 bp & -0.66\% & -0.084 & -17.83\% \\
1M & 5 bp & -3.00\% & -0.378 & -30.34\% \\
1M & 10 bp & -5.28\% & -0.664 & -41.16\% \\
1M & 15 bp & -7.51\% & -0.940 & -50.47\% \\
\midrule
3M & 0 bp & 4.41\% & 0.518 & -13.27\% \\
3M & 5 bp & 1.93\% & 0.226 & -22.64\% \\
3M & 10 bp & -0.50\% & -0.059 & -31.04\% \\
3M & 15 bp & -2.87\% & -0.337 & -38.57\% \\
\midrule
6M & 0 bp & -2.81\% & -0.401 & -28.50\% \\
6M & 5 bp & -4.43\% & -0.634 & -36.55\% \\
6M & 10 bp & -6.03\% & -0.862 & -44.14\% \\
6M & 15 bp & -7.60\% & -1.085 & -50.93\% \\
\bottomrule
\end{tabular}

\tablenote{The table reports transaction-cost sensitivity of the main Low-vol Gated Weekly Signal$\times$Risk strategy across re-estimation frequencies on the aligned common out-of-sample sample from 2015-03-30 to 2023-05-26 ($N=1947$). Annual return, Sharpe ratio, and maximum drawdown are reported under alternative one-way transaction-cost assumptions ranging from 0~bp to 15~bp.}
\end{table}

The transaction-cost analysis reinforces the previous frequency-robustness findings. Among the three specifications, the 3M strategy is the only frequency that remains economically viable under the baseline 5~bp transaction-cost assumption, preserving positive annual return and Sharpe ratio.

In contrast, both the 1M and 6M implementations exhibit negative net Sharpe ratios even at the baseline cost level, indicating either excessive turnover or insufficient predictive robustness. The 3M strategy's break-even transaction cost of approximately 8.96~bp further suggests that its profitability is economically meaningful but not highly cost-robust.

\begin{table}[H]
\centering
\footnotesize
\setlength{\tabcolsep}{4pt}
\renewcommand{\arraystretch}{1.05}
\caption{Reality Check and SPA tests across re-estimation frequencies}
\label{tab:frequency_rc_spa}

\begin{tabular}{lccccc}
\toprule
Test & Best Stat. & 95\% Crit. Value & p-value & Best Config. & No. Config. \\
\midrule
White Reality Check-style & 0.3522 & 1.7212 & 0.4636 & Strategy\_3M & 3 \\
Hansen SPA-style & 0.3522 & 1.5761 & 0.3948 & Strategy\_3M & 3 \\
\bottomrule
\end{tabular}

\tablenote{The tests compare the 1M, 3M, and 6M main strategies over the common out-of-sample period. The bootstrap uses 10,000 replications and a block length of 20 trading days.}
\end{table}

The Reality Check and SPA tests provide an important qualification. Although Strategy\_3M is identified as the strongest configuration based on the studentized mean differential relative to Buy-and-Hold, neither the White Reality Check-style test nor the Hansen SPA-style test rejects the data-snooping null hypothesis. The corresponding p-values are 0.4636 and 0.3948, respectively.

Therefore, the 3M specification should not be interpreted as statistically dominant after multiple-testing adjustment. Instead, it is retained as the baseline frequency because it combines the strongest HAC-based return-prediction evidence with the most economically viable main-strategy performance under realistic transaction costs.

\subsection{Robustness to the XGBoost hyperparameter grid}
\label{subsec:xgb_grid_robustness}

One potential concern regarding the return-forecasting stage is that the baseline XGBoost specification uses a deliberately conservative hyperparameter grid, which may limit model flexibility and potentially understate predictive performance. To address this concern, an additional robustness exercise is conducted using an expanded hyperparameter grid.

The expanded grid varies the number of boosting trees, tree depth, and learning rate according to
\begin{equation}
\begin{aligned}
n_{\text{estimators}}
&\in \{100,300,500\},\\
\text{max\_depth}
&\in \{2,3,4\},\\
\eta
&\in \{0.01,0.03,0.05\}.
\end{aligned}
\end{equation}

All remaining regularization parameters are held fixed or kept close to their baseline values. This design generates 27 candidate configurations at each walk-forward re-estimation date, compared with the more restrictive baseline grid.

The expanded-grid experiment is implemented under the same walk-forward framework as the baseline return-prediction model. At each re-estimation anchor, model selection is based exclusively on validation-sample correlation between predicted and realized returns. The selected model is then used to generate strictly out-of-sample forecasts. Consequently, the experiment preserves the no-look-ahead structure of the baseline framework.

The empirical evidence does not support replacing the baseline conservative grid with the expanded specification. Over the feasible out-of-sample period from 2014-12-17 to 2023-05-26, the expanded-grid model achieves an out-of-sample correlation of 0.0124 and a directional accuracy of 52.25\%, both lower than those obtained under the baseline specification. Forecast-error measures also deteriorate modestly, with MSE increasing to \(2.3950 \times 10^{-4}\) and MAE increasing to 0.01027.

These results suggest that additional model flexibility does not translate into superior out-of-sample predictive performance. A plausible interpretation is that CSI 300 daily return predictability is weak, unstable, and characterized by a low signal-to-noise ratio. Under such conditions, a wider hyperparameter grid may fit transient validation-sample noise rather than identify robust predictive structure.

Accordingly, the baseline conservative grid is retained as the main specification for the return-prediction framework. The expanded-grid exercise is interpreted as a robustness and limitation check rather than as an alternative preferred model.

\subsection{Early- versus later-period return prediction performance}
\label{subsec:early_late_oos}

One potential concern regarding the return-prediction framework is that the initial walk-forward estimation windows rely on relatively limited training data. As a result, early out-of-sample forecasts may be more sensitive to estimation uncertainty than forecasts generated later in the sample, when substantially more historical data are available.

To evaluate this possibility, the return-prediction out-of-sample period is divided into an early phase and a later phase. This diagnostic is not used for model selection, but is intended to assess whether predictive performance changes materially as the expanding estimation sample grows over time.

\begin{table}[H]
\centering
\footnotesize
\setlength{\tabcolsep}{4pt}
\renewcommand{\arraystretch}{1.05}
\caption{Early- versus later-period return prediction performance}
\label{tab:early_late_return_prediction}

\begin{tabular}{lcccccc}
\toprule
Period & Dates & Obs & Corr. & MAE & Hit Ratio & HAC p-value \\
\midrule
Early OOS & 2014-12-17--2016-12-31 & 489 & 0.0687 & 0.0145 & 52.35\% & 0.2882 \\
Later OOS & 2017-01-01--2023-05-26 & 1515 & 0.0040 & 0.0088 & 53.86\% & 0.0014 \\
\bottomrule
\end{tabular}

\tablenote{Early OOS corresponds to the initial phase of the expanding-window framework, while Later OOS reflects a substantially larger estimation history. Hit Ratio denotes the fraction of observations for which predicted and realized returns have the same sign. HAC p-values are based on Newey--West standard errors for the null hypothesis that directional accuracy equals 50\%.}
\end{table}

Table~\ref{tab:early_late_return_prediction} reports the results. The early out-of-sample period exhibits a higher predictive correlation, but also substantially higher forecast error, reflecting the turbulent market conditions associated with the 2015--2016 Chinese equity-market crash. Despite the stronger raw correlation, directional accuracy is not statistically significant under HAC inference.

In contrast, the later out-of-sample period produces a much lower predictive correlation, but a significantly stronger directional-accuracy result. The hit ratio rises to 53.86\%, with a HAC-adjusted p-value of 0.0014, while forecast errors decline materially.

These findings suggest that predictive usefulness becomes more stable as the expanding training sample matures. Although the later sample exhibits weaker linear association between predicted and realized returns, it generates more reliable sign prediction, which is arguably more relevant for practical trading implementation.

\subsection{Feature-ablation analysis}
\label{subsec:feature_ablation}

This subsection evaluates which feature groups contribute most strongly to the economic performance of the final Low-vol Gated Weekly Signal$\times$Risk strategy. All ablation models are aligned to the main out-of-sample period from 2014-12-17 to 2023-05-26, comprising 2,004 observations.

The ablation settings remove specific groups of predictors as follows:
\begin{itemize}
\item \textbf{No Volatility Forecast}: excludes the forecasted volatility predictor, denoted by \texttt{logRVhat\_t1};
\item \textbf{No Regime}: excludes all regime-related predictors, including the filtered regime probability $p_t$, its lagged value (\texttt{pt\_lag5}), and the volatility--regime interaction term (\texttt{vol\_pt});
\item \textbf{No High-Order Risk}: excludes higher-moment risk measures, namely $\log(RQ_t)$ and the signed jump measure (\texttt{signed\_jump});
\item \textbf{No Forecast/Regime Conditioning}: excludes \texttt{logRVhat\_t1}, $p_t$, \texttt{pt\_lag5}, \texttt{vol\_pt}, \texttt{vol\_lag5}, and \texttt{vol\_lag22}.
\end{itemize}

The same Low-vol Gated Weekly implementation is then applied to each specification, ensuring that observed performance differences reflect feature-content variation rather than changes in strategy design.

\begin{table}[H]
\centering
\footnotesize
\setlength{\tabcolsep}{4pt}
\renewcommand{\arraystretch}{1.05}
\caption{Feature ablation: Economic performance of the Low-vol Gated Weekly strategy}
\label{tab:feature_ablation_econ}

\begin{tabular}{lccccc}
\toprule
Model & Ann. Return & Ann. Vol & Sharpe & MaxDD & Break-even Cost \\
\midrule
Main Model & 1.93\% & 7.56\% & 0.2551 & -14.52\% & 19.97 bp \\
No Vol Forecast & 0.73\% & 8.14\% & 0.0902 & -20.90\% & 10.83 bp \\
No Regime & 1.13\% & 6.27\% & 0.1800 & -17.60\% & 15.88 bp \\
No High-Order Risk & 1.93\% & 7.42\% & 0.2604 & -14.39\% & 20.56 bp \\
No Forecast/Regime Conditioning & -0.06\% & 6.90\% & -0.0083 & -19.22\% & 4.59 bp \\
\bottomrule
\end{tabular}

\tablenote{The table reports 5~bp net performance for feature-ablation strategies over the aligned main out-of-sample sample from 2014-12-17 to 2023-05-26 ($N=2004$). Break-even cost denotes the transaction-cost level at which the strategy Sharpe ratio falls to zero.}
\end{table}

The ablation results show that the full model delivers the strongest overall balance between return, risk control, drawdown management, and transaction-cost robustness.

Removing the volatility forecast materially reduces performance. Sharpe declines from 0.2551 to 0.0902, while maximum drawdown deteriorates from -14.52\% to -20.90\%. Removing regime-related features also weakens performance, with Sharpe declining to 0.1800 and maximum drawdown increasing to -17.60\%. These results support the interpretation that volatility forecasts and regime information contribute economically by improving risk scaling, signal reliability, and exposure conditioning.

The No High-Order Risk specification performs similarly to the main model and even produces a slightly higher Sharpe ratio in this implementation. This suggests that higher-order risk features are not essential for the final strategy's economic performance, although they may still contribute to predictive structure in other diagnostics.

By contrast, removing forecast and regime-conditioning components causes the strategy to lose nearly all economic value. The resulting Sharpe ratio falls to approximately zero and the break-even transaction cost declines substantially.

\begin{table}[htbp]
\centering
\footnotesize
\setlength{\tabcolsep}{4pt}
\renewcommand{\arraystretch}{1.05}
\caption{Feature ablation: Bootstrap Sharpe difference tests}
\label{tab:feature_ablation_bootstrap}

\begin{tabular}{lccc}
\toprule
Comparison & Sharpe Diff. & 95\% Bootstrap CI & p-value ($A>B$) \\
\midrule
Main vs Buy \& Hold & 0.3024 & [-0.6699, 1.1652] & 0.2760 \\
Main vs No Vol Forecast & 0.1649 & [-0.3553, 0.6461] & 0.2752 \\
Main vs No Regime & 0.0751 & [-0.5401, 0.6385] & 0.4123 \\
Main vs No High-Order Risk & -0.0053 & [-0.5190, 0.4973] & 0.5091 \\
Main vs No Forecast/Regime Conditioning & 0.2634 & [-0.4836, 1.0153] & 0.2362 \\
\bottomrule
\end{tabular}

\tablenote{The table reports circular block bootstrap Sharpe-difference tests at 5~bp transaction costs. Positive Sharpe differences indicate stronger performance for the main model relative to the comparison strategy.}
\end{table}

The bootstrap evidence supports a cautious interpretation of the ablation results. Although the main model generally exhibits higher observed Sharpe ratios than most ablation variants, none of the Sharpe differences are statistically significant at conventional confidence levels.

Therefore, the ablation analysis should be interpreted primarily as economic sensitivity evidence rather than definitive statistical proof of feature dominance.

\subsection{Strategy decomposition}
\label{subsec:strategy_decomposition}

To further clarify the economic sources of performance, this subsection decomposes the baseline Low-vol Gated Weekly Signal$\times$Risk strategy into its core value components. The objective is to distinguish whether economic performance primarily arises from directional return forecasting, volatility-based risk scaling, or their interaction within the final implementation framework.

Three decomposition variants are considered:
\begin{itemize}
\item a \textit{Combined Signal$\times$Risk} strategy that jointly uses predicted returns and forecasted volatility;
\item a \textit{Return-only} strategy that isolates directional forecasting;
\item a \textit{Volatility-only} strategy that isolates volatility-based position sizing.
\end{itemize}

All decomposition results use the identical baseline 3M re-estimation frequency, the same walk-forward calibration framework, the same low-volatility gating rule ($p_t \leq 0.5$), weekly rebalancing, threshold filtering, position caps, and no-trade-band controls. The common out-of-sample evaluation period is 2014-12-17 to 2023-05-26 ($N=2004$).

Let $\hat r_{t+1}$ denote the XGBoost-predicted next-period return, and let
\[
\hat \sigma_{t+1}
=
\sqrt{\exp(\widehat{\log RV}_{t+1})}
\]
denote forecasted next-period volatility derived from the HARQ-based volatility model.

The three raw decomposition signals are defined as
\begin{equation}
\begin{aligned}
s_t^{\text{comb}}
&=
\frac{\hat r_{t+1}}
{\hat \sigma_{t+1}},
\\
s_t^{\text{ret}}
&=
\hat r_{t+1},
\\
s_t^{\text{vol}}
&=
\frac{1}
{\hat \sigma_{t+1}}.
\end{aligned}
\end{equation}

The Combined specification corresponds directly to the main Signal$\times$Risk framework, where return forecasts are volatility scaled. The Return-only specification removes volatility scaling entirely, isolating pure directional prediction. The Volatility-only specification removes return forecasts and allocates exposure inversely to predicted volatility.

All three specifications are then passed through the identical walk-forward implementation pipeline, including low-volatility gating, rolling threshold calibration, exposure scaling, position caps, weekly rebalancing, and no-trade-band controls.

An important implementation distinction is that the Volatility-only specification is long-only by construction in the empirical implementation, whereas the Combined and Return-only specifications remain unrestricted and may take both long and short positions depending on the sign of the return forecast.

To evaluate whether decomposition results are primarily driven by short-selling flexibility, an additional long-only decomposition is considered. Under this constraint, negative predicted returns are truncated at zero:
\begin{equation}
\begin{aligned}
s_t^{\text{comb,LO}}
&=
\frac{\max(\hat r_{t+1},0)}
{\hat \sigma_{t+1}},
\\
s_t^{\text{ret,LO}}
&=
\max(\hat r_{t+1},0),
\\
s_t^{\text{vol,LO}}
&=
\frac{1}
{\hat \sigma_{t+1}}.
\end{aligned}
\end{equation}

This extension allows direct comparison between unrestricted and implementation-constrained economic value.

\begin{table}[H]
\centering
\footnotesize
\setlength{\tabcolsep}{4pt}
\renewcommand{\arraystretch}{1.05}
\caption{Low-vol Gated strategy decomposition}
\label{tab:strategy_decomposition}

\begin{tabular}{lccccc}
\toprule
Strategy & Cost & Ann. Return & Ann. Vol & Sharpe & MaxDD \\
\midrule
Combined Signal$\times$Risk & 0 bp & 2.58\% & 7.56\% & 0.341 & -14.17\% \\
Return-only & 0 bp & 1.27\% & 7.62\% & 0.166 & -15.30\% \\
Volatility-only & 0 bp & -0.14\% & 6.17\% & -0.022 & -14.19\% \\
\midrule
Combined Signal$\times$Risk & 5 bp & 1.93\% & 7.56\% & 0.255 & -14.52\% \\
Return-only & 5 bp & 0.66\% & 7.62\% & 0.086 & -18.07\% \\
Volatility-only & 5 bp & -0.45\% & 6.17\% & -0.073 & -15.16\% \\
Buy-and-Hold & 5 bp & -1.11\% & 23.38\% & -0.047 & -50.60\% \\
\bottomrule
\end{tabular}

\tablenote{The table decomposes the final Low-vol Gated Weekly Signal$\times$Risk strategy over the main 3M out-of-sample sample from 2014-12-17 to 2023-05-26 ($N=2004$) into combined, return-only, and volatility-only components.}
\end{table}

The decomposition results show that the Combined Signal$\times$Risk specification performs best. At 5~bp transaction costs, it achieves a Sharpe ratio of 0.255, compared with 0.086 for the Return-only component and -0.073 for the Volatility-only component.

This evidence indicates that the strategy's economic value does not arise from volatility scaling alone or from return forecasting alone. Instead, performance is generated by combining directional forecasts with volatility-based risk scaling and low-volatility gating.

\begin{table}[H]
\centering
\footnotesize
\setlength{\tabcolsep}{4pt}
\renewcommand{\arraystretch}{1.05}
\caption{Long-only strategy decomposition}
\label{tab:strategy_decomposition_longonly}

\begin{tabular}{lccccc}
\toprule
Strategy & Cost & Ann. Return & Ann. Vol & Sharpe & MaxDD \\
\midrule
Combined Long-only Signal$\times$Risk & 0 bp & 0.14\% & 5.41\% & 0.026 & -14.30\% \\
Return-only Long-only & 0 bp & -0.25\% & 5.25\% & -0.047 & -16.01\% \\
Volatility-only Long-only & 0 bp & -0.14\% & 6.17\% & -0.022 & -14.19\% \\
Buy-and-Hold & 0 bp & -1.10\% & 23.38\% & -0.047 & -50.60\% \\
\midrule
Combined Long-only Signal$\times$Risk & 5 bp & -0.25\% & 5.41\% & -0.046 & -15.06\% \\
Return-only Long-only & 5 bp & -0.60\% & 5.25\% & -0.115 & -16.80\% \\
Volatility-only Long-only & 5 bp & -0.45\% & 6.17\% & -0.073 & -15.16\% \\
Buy-and-Hold & 5 bp & -1.11\% & 23.38\% & -0.047 & -50.60\% \\
\bottomrule
\end{tabular}

\tablenote{The table reports long-only decomposition results over the main out-of-sample sample from 2014-12-17 to 2023-05-26 ($N=2004$).}
\end{table}

The long-only decomposition reveals an important implementation limitation. Before transaction costs, the Combined Long-only Signal$\times$Risk strategy achieves only a modest Sharpe ratio of 0.026. After 5~bp transaction costs, the Sharpe ratio declines to -0.046, almost identical to the Buy-and-Hold Sharpe ratio of -0.047.

Removing short positions therefore eliminates most of the economic value generated by the unrestricted strategy. Although the framework remains defensive under long-only constraints, reducing annualized volatility from 23.38\% to 5.41\% and maximum drawdown from -50.60\% to -15.06\%, it no longer generates economically meaningful positive net returns.

\subsection{Implementation robustness: Normalized naive Signal$\times$Risk benchmark}
\label{subsec:naive_signal_risk}

To evaluate whether the final strategy's signal-engineering components are necessary, this subsection compares the final implementation with a normalized naive Signal$\times$Risk benchmark.

The benchmark uses the raw ratio of predicted return to predicted volatility,
\begin{equation}
w_t
=
c_t
\cdot
\frac{\hat r_{t+1}}
{\hat \sigma_{t+1}},
\end{equation}
subject to
\begin{equation}
|w_t|
\leq
w_{\max}.
\end{equation}

The scaling factor $c_t$ is estimated using only information available prior to date $t$, ensuring that the benchmark remains free of look-ahead bias. Unlike the final strategy, the benchmark does not use low-volatility gating, threshold filtering, weekly rebalancing, or no-trade-band controls.

\begin{table}[H]
\centering
\footnotesize
\setlength{\tabcolsep}{5pt}
\renewcommand{\arraystretch}{1.05}
\caption{Normalized naive Signal$\times$Risk benchmark}
\label{tab:naive_signal_risk}

\begin{tabular}{lccccc}
\toprule
Strategy & Ann. Return & Ann. Vol & Sharpe & MaxDD & Turnover \\
\midrule
Buy-and-Hold (net) & -1.11\% & 23.38\% & -0.047 & -50.60\% & 0.0005 \\
Normalized Naive Signal$\times$Risk (gross) & 5.73\% & 12.90\% & 0.444 & -38.46\% & 0.2649 \\
Normalized Naive Signal$\times$Risk (net) & 2.26\% & 12.88\% & 0.176 & -46.85\% & 0.2649 \\
\bottomrule
\end{tabular}

\tablenote{The normalized naive Signal$\times$Risk strategy uses the raw ratio of predicted return to predicted volatility and applies an expanding-window scaling procedure using only past information.}
\end{table}

The normalized naive benchmark confirms that the raw Signal$\times$Risk signal contains economically useful information. It generates a positive gross annualized return of 5.73\% and remains profitable after transaction costs, with a net annualized return of 2.26\% and a net Sharpe ratio of 0.176.

However, the benchmark remains costly and insufficiently robust as a standalone implementation. Average turnover reaches 0.2649, substantially higher than that of the final strategy. Transaction costs reduce the Sharpe ratio from 0.444 to 0.176, while maximum drawdown remains elevated at -46.85\%.

These findings support the final implementation design. The raw predictive signal is informative, but without low-volatility gating, threshold filtering, weekly rebalancing, and no-trade-band controls, much of its economic value is eroded by turnover and drawdown exposure.

\subsection{Strategy parameter sensitivity and data-snooping adjustment}
\label{subsec:qb_sensitivity_rc_spa}

This subsection evaluates whether the main strategy results depend on the choice of threshold quantile $q$ and no-trade-band parameter $b$.

The strategy is evaluated across the parameter grid with
$q \in \{0.50, 0.55, 0.60, 0.65, 0.70\}$
and
$b \in \{0.00, 0.01, 0.02, 0.03\}$.

The strongest net Sharpe ratio is obtained around $q=0.65$ and $b=0.01$ or $b=0.02$, reaching approximately 0.269. The conservative baseline specification, $q=0.60$ and $b=0.02$, remains close to the best-performing region, with a net Sharpe ratio of approximately 0.255 and a favorable drawdown profile.

\begin{figure}[H]
\centering
\caption{$q \times b$ sensitivity of net Sharpe}
\label{fig:qb_sensitivity}
\includegraphics[width=0.85\textwidth]{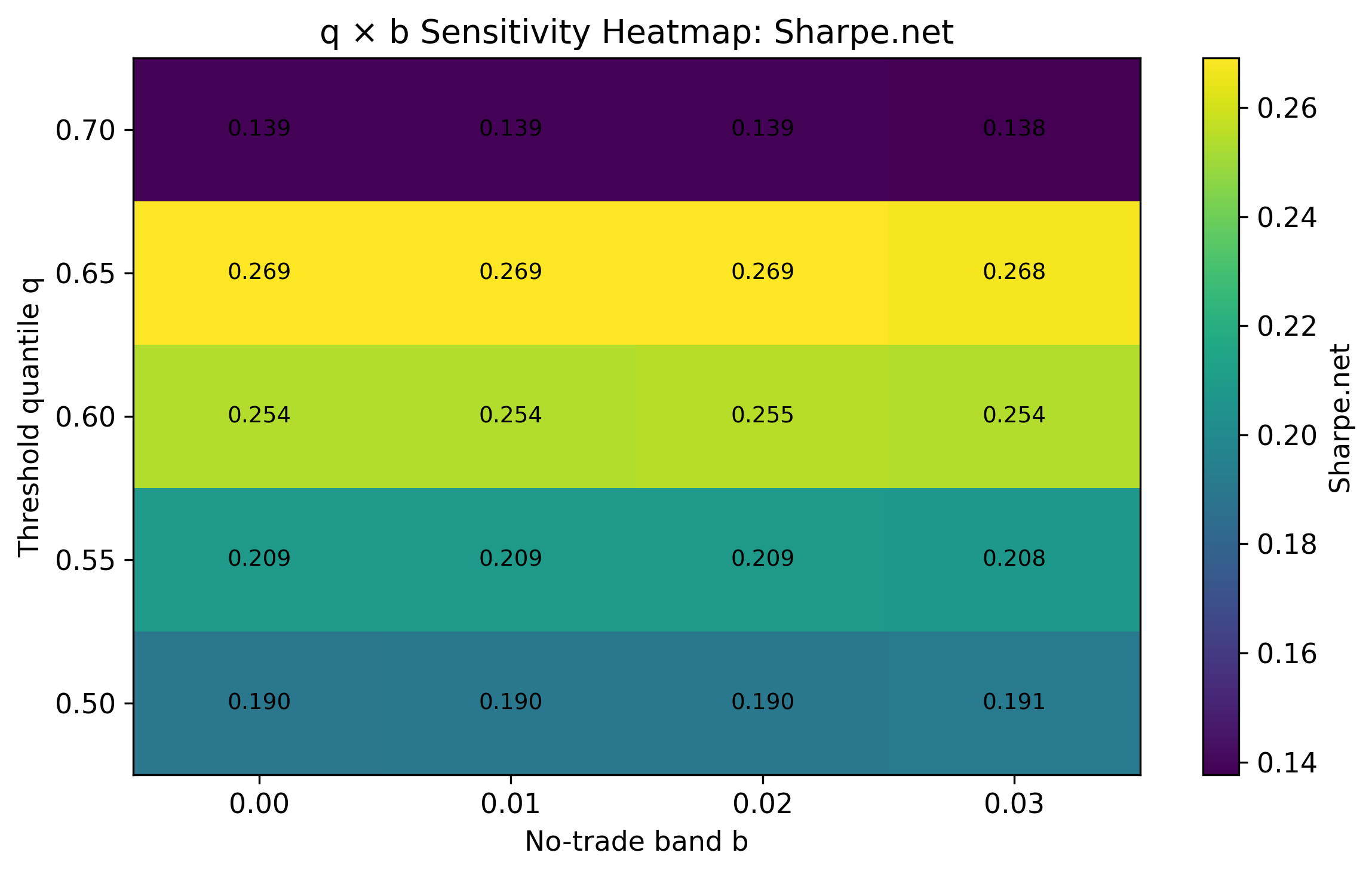}

\figurenote{The figure reports net Sharpe ratios across threshold quantiles $q$ and no-trade bands $b$. The baseline specification $q=0.60$, $b=0.02$ is retained as a conservative ex ante choice rather than selected as the ex post best-performing parameter combination.}
\end{figure}

White Reality Check-style and Hansen SPA-style procedures are applied across the entire $q \times b$ grid. The resulting p-values are approximately 0.502 and 0.495, respectively. Consequently, the apparent superiority of the best-performing parameter combination is not statistically significant after accounting for data snooping and repeated parameter selection.

The parameter-sensitivity analysis should therefore be interpreted as evidence of local stability rather than as justification for ex post optimization. Performance remains reasonably stable throughout the region surrounding the baseline specification, particularly for $q$ values between 0.60 and 0.65.

Accordingly, the baseline specification is retained as a conservative implementation choice rather than replaced by the ex post best-performing parameter combination.

\subsection{Summary of robustness analyses}
\label{subsec:robustness_summary}

The robustness analyses support a cautious but coherent interpretation of the study findings.

First, volatility forecasting is robust across alternative regime specifications and re-estimation frequencies. Regime-augmented HARQ models consistently outperform the baseline HARQ model, while Mincer--Zarnowitz regressions and conditional Diebold--Mariano tests indicate that the gains are particularly relevant during elevated-volatility periods. Differences between hard and soft regime representations are relatively small, suggesting that the primary benefit arises from capturing regime dependence itself rather than from a specific transformation of the regime variable.

Second, return prediction is more sensitive to re-estimation frequency than volatility forecasting. On the common out-of-sample sample, the 3M specification is the only frequency that produces statistically significant HAC-corrected directional accuracy in both the full sample and the low-volatility regime. It is also the only main Signal$\times$Risk implementation that remains economically viable under 5~bp transaction costs. However, Reality Check and SPA procedures do not establish the 3M specification as statistically dominant after accounting for multiple-testing effects. Consequently, frequency selection should be interpreted cautiously and viewed as economically suggestive rather than statistically definitive.

Third, benchmark comparisons demonstrate that simpler strategies, including momentum and regime-only rules, can perform well under certain conditions. Nevertheless, these alternatives exhibit weaker stability across frequencies and do not replace the central mechanism proposed in this study. The main strategy is best viewed as a disciplined integration of return forecasts, volatility scaling, and regime conditioning rather than as a simple market-timing rule.

Fourth, the feature-ablation and decomposition analyses indicate that the economic value of the final strategy arises from the interaction between return forecasts, volatility scaling, regime conditioning, and implementation design. Removing volatility forecasts, regime variables, or forecast/regime-conditioning components weakens performance, although bootstrap inference suggests that these differences should be interpreted as economic sensitivity evidence rather than statistically decisive proof of feature dominance.

Fifth, the normalized naive Signal$\times$Risk benchmark confirms that the raw predictive signal contains economically useful information but is not practically robust without signal engineering. Low-volatility gating, threshold filtering, weekly rebalancing, and no-trade-band controls play an important role in reducing turnover, drawdowns, and implementation fragility.

Overall, the robustness analyses reinforce the central conclusion of the study. The proposed framework does not uncover strong unconditional return predictability. Instead, it shows that weak and state-dependent predictive signals can be transformed into economically meaningful defensive performance when combined with volatility information, regime filtering, and disciplined implementation rules. At the same time, the statistical evidence remains modest. Consequently, the final interpretation emphasizes economic usefulness, risk management, and implementation robustness rather than strong statistical alpha.

\section{Conclusion}
\label{sec:conclusion}

This study examines whether regime-dependent volatility modeling and machine-learning-based return prediction can jointly improve statistical forecasting performance and economic strategy outcomes in the Chinese equity market. Using CSI 300 high-frequency data within a strictly walk-forward framework, the analysis integrates a sequential two-stage architecture. Stage 1 models realized volatility through regime-augmented HARQ specifications combined with MS--GJR--GARCH filtering, while Stage 2 employs XGBoost return prediction using volatility, regime, and return-related predictors. The broader objective is not merely to evaluate predictive accuracy, but to assess whether weak state-dependent signals can be transformed into economically relevant performance under realistic implementation constraints.

The empirical evidence supports the first-stage volatility-forecasting hypothesis. Across alternative model specifications and re-estimation frequencies, regime-augmented HARQ models consistently outperform baseline HARQ benchmarks under QLIKE, MSE, and LMAE evaluation criteria. HAC-robust Diebold--Mariano tests provide statistical support for these forecasting gains. Mincer--Zarnowitz regressions and conditional regime diagnostics further suggest that regime dependence improves volatility-forecasting robustness, particularly during elevated-volatility periods. Importantly, soft filtered probabilities and hard regime indicators generate broadly similar improvements, implying that the primary benefit arises from incorporating regime information itself rather than from a particular regime functional form.

In contrast, the second-stage return-prediction evidence is substantially weaker and more conditional. Return predictability remains fragile, economically modest, and concentrated primarily in low-volatility regimes. HAC-adjusted hit-ratio tests, rolling information-coefficient diagnostics, and early-versus-later out-of-sample comparisons collectively indicate that predictive content is neither stable nor unconditional. Although the 3M specification provides the most statistically defensible return-prediction evidence within the main framework, predictive performance remains sensitive to model design, re-estimation frequency, and implementation assumptions. High-volatility states are consistently characterized by weak or negative directional forecasting ability, implying that the primary role of regime awareness in return prediction is often exposure suppression rather than directional alpha generation.

The results further demonstrate that implementation design is critical. Raw predictive signals and naive Signal$\times$Risk frameworks generally lose much of their economic value after realistic trading frictions are introduced. However, implementation choices including low-volatility gating, walk-forward threshold calibration, volatility scaling, weekly rebalancing, and turnover controls materially improve defensive performance. At the same time, these gains should be interpreted cautiously. Bootstrap Sharpe-ratio tests, Reality Check procedures, and Hansen SPA tests indicate that although selected implementations may generate economically attractive outcomes, statistical dominance over alternative configurations remains limited after correcting for configuration-space data snooping. Consequently, the results are better characterized as economically suggestive evidence of defensive implementation design rather than definitive evidence of alpha discovery.

Feature-ablation analyses, regime-only benchmarks, and strategy decomposition further refine this interpretation. Volatility and regime information contribute primarily through risk control, exposure timing, and implementation stability, whereas the incremental contribution of machine-learning return forecasts is more limited than initially expected. In particular, long-short implementations retain most of the framework's economic potential, whereas the long-only decomposition shows that much of the predictive edge weakens materially once realistic directional constraints are imposed. These findings do not invalidate the framework, but they clarify that its practical usefulness depends heavily on the implementation vehicle and market-access conditions.

The study contributes in four principal ways. First, it develops and evaluates an integrated regime-aware volatility-plus-return architecture for the CSI 300 market, extending volatility-regime research into a broader predictive and economic framework. Second, it emphasizes strict walk-forward implementation, filtered-probability discipline, and extensive robustness diagnostics, thereby improving methodological transparency relative to simpler predictive studies. Third, it demonstrates that economic value depends more strongly on signal engineering and implementation discipline than on raw predictive strength alone. Fourth, it provides evidence that volatility and regime information are substantially more robust than direct return forecasting, suggesting that future quantitative frameworks may benefit more from defensive state conditioning than from aggressive directional prediction.

Several limitations remain. First, the framework focuses on a single market and may not generalize directly across asset classes or international settings. Second, transaction-cost assumptions, short-selling feasibility, and implementation frictions remain simplified. This limitation is particularly important because practical short-selling opportunities in Chinese equity markets may be materially more constrained than assumed in the empirical analysis. Third, generated-regressor dependence, arising from the use of volatility forecasts as inputs into the return-prediction model, introduces additional model uncertainty. Fourth, although XGBoost serves as a useful nonlinear benchmark, alternative machine-learning frameworks may generate different robustness-versus-overfitting trade-offs. Finally, while multiple-testing corrections are expanded relative to many empirical strategy studies, strategy-selection uncertainty cannot be eliminated completely.

Future research should therefore focus on three directions. First, applying the framework to alternative implementation vehicles, such as CSI 300 futures, ETFs, and broader cross-market settings, would improve external validity. Second, jointly modeling volatility regimes and return predictability within a unified architecture may better capture endogenous state dependence than the current sequential framework. Third, future work should further enhance implementation realism through richer transaction-cost modeling, explicit short-sale constraints, adaptive parameterization, and broader benchmark integration. More broadly, the findings suggest that the next frontier may lie less in discovering strong unconditional return predictability and more in designing robust defensive allocation systems that exploit weak but persistent state dependence.

Overall, this study concludes that unconditional return predictability remains weak, unstable, and highly implementation sensitive. Nevertheless, regime-aware volatility modeling combined with disciplined signal engineering can improve defensive economic outcomes through exposure control, volatility conditioning, and selective participation. The central implication is therefore not that machine learning delivers strong unconditional return forecasts, but that carefully structured regime-volatility frameworks may help transform weak conditional information into more resilient portfolio designs under specific implementation environments.

\section*{Data and code availability}
\addcontentsline{toc}{section}{Data and code availability}

The replication files required to reproduce the empirical results reported in this paper, including data-processing scripts, volatility-model estimation procedures, return-prediction models, and strategy backtesting code, are available in the accompanying GitHub repository.

The CSI 300 high-frequency data used in this study are obtained from commercial databases and therefore cannot be redistributed publicly. Processed data files necessary for replication are provided where permitted by the data license.

\bibliographystyle{elsarticle-harv}
\bibliography{references}

\end{document}